\documentclass[a4paper, 12pt]{article}  %
\usepackage[top=1in, bottom=1in, left=1in, right=1in]{geometry}
\usepackage{graphicx}
\usepackage{amsthm}
\usepackage{amsmath}
\usepackage{amsfonts}
\usepackage{amssymb}
\usepackage{mathtools}
\usepackage{pdfpages}
\usepackage{hyperref}
\usepackage{natbib}
\usepackage[english]{babel}
\usepackage{txfonts}
\usepackage{caption}
\usepackage[figuresright]{rotating}
\usepackage{listings,relsize}
\usepackage{setspace}
\usepackage{color}
\usepackage{tcolorbox}
\usepackage{chngcntr}
\usepackage[normalem]{ulem}
\usepackage{cancel}
\usepackage{lineno}
\usepackage{soul}
\usepackage{setspace}
\usepackage{longtable}
\usepackage{tabularx}
\usepackage{booktabs}
\usepackage{array}
\usepackage{ragged2e}

\newcommand{\thinvert}{\,\vert\,}

\begin{document}

%[Adaptive Smoothing for Phylodynamics: Web Appendices]
\title{\large{\textbf{Horseshoe-based Bayesian nonparametric estimation of effective population size trajectories} } } 

\author{ \small{\textbf{James R. Faulkner}$^{1,2}$, \textbf{Andrew F. Magee}$^3$, \textbf{Beth Shapiro}$^{4,5}$, \textbf{Vladimir N. Minin}$^{6,*}$ } \\
\footnotesize{ $^1$Quantitative Ecology and Resource Management, University of Washington, Seattle, WA 98195, U.S.A.}\\
\footnotesize{ $^2$National Marine Fisheries Service, Northwest Fisheries Science Center, Seattle, WA 98112, U.S.A.}\\
%\emailx{jim.faulkner@noaa.gov}
\footnotesize{ $^3$Department of Biology, University of Washington, Seattle, WA 98195, U.S.A.}\\
\footnotesize{ $^4$Ecology and Evolutionary Biology Department and Genomics Institute,} \\
\footnotesize{ University of California Santa Cruz, Santa Cruz, CA 95064, U.S.A.}\\
\footnotesize{ $^5$Howard Hughes Medical Institute, University of California Santa Cruz, Santa Cruz, CA 95064, U.S.A.}\\
\footnotesize{ $^{6}$Department of Statistics, University of California Irvine, Irvine, CA 92697, U.S.A.}\\
\footnotesize{ $^{*}$Corresponding author: {\tt vminin@uci.edu} } }

%\vspace{-1cm}

\date{}

%\volume{XX}
%\pubyear{2019}
%\artmonth{April}
%\doi{10x}

\maketitle

%\label{firstpage}

\begin{abstract}
Phylodynamics is an area of population genetics that uses genetic sequence data to estimate past population dynamics.
Modern state-of-the-art Bayesian nonparametric methods for recovering population size trajectories of unknown form use either change-point models or Gaussian process priors.
Change-point models suffer from computational issues when the number of change-points is unknown and needs to be estimated.
Gaussian process-based methods lack local adaptivity and cannot accurately recover trajectories that exhibit features such as abrupt changes in trend or varying levels of smoothness.
We propose a novel, locally-adaptive approach to Bayesian nonparametric phylodynamic inference that has the flexibility to accommodate a large class of functional behaviors.
Local adaptivity results from modeling the log-transformed effective population size \emph{a priori} as a horseshoe Markov random field, a recently proposed statistical model that blends together the best properties of the change-point and Gaussian process modeling paradigms.
We use simulated data to assess model performance, and find that our proposed method results in reduced bias and increased precision when compared to contemporary methods.
We also use our models to reconstruct past changes in genetic diversity of human hepatitis C virus in Egypt and to estimate population size changes of ancient and modern steppe bison.
These analyses show that our new method captures features of the population size trajectories that were missed by the state-of-the-art methods.  \\
\end{abstract}


\section{Introduction}
\label{s:intro}
Estimation of population sizes and population dynamics over time is an important task in ecology and epidemiology.
Census population sizes can be difficult to estimate due to infeasible sampling requirements or study costs.
Genetic sequences are a growing source of information that can be used to infer past population sizes from the signatures of genetic diversity.
Phylodynamics is a discipline that uses genetic sequence data to estimate past population dynamics.
Many phylodynamic models draw on coalescent theory \citep{kingman1982, griffiths1994}, which provides a probabilistic framework that connects the branching times of a genealogical tree with the effective population size and other demographic variables, such as migration rates, of the population from which the genealogy was drawn.
Effective population size can be interpreted as a measure of genetic diversity in a population and is proportional to census population size if coalescent model assumptions are met.
When genetic diversity is high, the effective population size approaches the census population size, given random mating and no inbreeding or genetic drift, but is otherwise smaller than the census size.
In our work we concentrate on estimation of effective population sizes over evolutionary time, which can be short for rapidly evolving virus populations and longer (but still estimable with preserved ancient molecular sequence samples) for more slowly-evolving organisms.
Some examples of successful application of phylodynamics include describing seasonal trends of influenza virus spread around the world \citep{rambaut2008}, quantifying dynamics of outbreaks like hepatitis C \citep{pybus2003} and Ebola viruses \citep{alizon2014}, and assessing the effects of climate change on populations of large mammals during the ice ages using ancient DNA \citep{shapiro2004, lorenzen2011}.
\par
Some approaches to phylodynamics use parametric functional relationships to describe effective population size trajectories \citep[\emph{e.g.,}][]{pybus2003, rasmussen2014}, but nonparametric methods offer a flexible alternative when an accurate estimate of a complex population size trajectory is needed and knowledge of the mechanisms driving population size changes is incomplete.
Nonparametric models have a long history of use in inferring effective population size trajectories.
 \cite{pybus2000} introduced a nonparametric method, called the skyline plot, that produced point-wise estimates of population size, where the number of estimates was equal to the number of sampled genetic sequences minus one.
The estimates from this method were highly variable, so a modification, referred to as the generalized skyline plot, created a set of discrete time interval groups that shared a single effective population size \citep{strimmer2001}.
These likelihood-based approaches were adapted to a Bayesian framework with the Bayesian skyline plot \citep{drummond2005} and the variable-knot spline approach of \cite{opgen2005}.
\cite{minin2008sky} provided an alternative to these change-point methods by introducing a Gaussian Markov random field (GMRF) smoothing prior that connected the piecewise-constant population size estimates between coalescent events without needing to specify or estimate knot locations.
\cite{palacios2012} and \cite{gill2013} extended the GMRF approach of \cite{minin2008sky} by constructing a GMRF prior on a discrete uniform grid.
A grid-free approach, introduced by \cite{palacios2013}, allowed the population size trajectories to vary continuously by using a Gaussian process (GP) prior.
\par
Modern nonparametric Bayesian methods offer the state-of-the-art for recovering effective population size trajectories of unknown form.
However, current methods cannot accurately recover trajectories that exhibit challenging features such as abrupt changes or varying levels of smoothness.
Such features may arise in populations in the form of bottlenecks, rapid population changes, or aperiodic fluctuations with varying amplitudes.
Accurate estimation of features like these can be important for understanding the demographic history of a population.
Outside of phylodynamics, various nonparametric statistical methods have been developed to deal with such nonstationary or locally-varying behavior under more standard likelihoods.
These methods include, but are not limited to, GPs with nonstationary covariance functions \citep{paciorek2006},  nonstationary process convolutions \citep{higdon1998, fuentes2002}, non-Gaussian Mat\'ern fields \citep{wallin2015}, and adaptive smoothing splines \citep{yue2012, yue2014}.
Each of these methods has good qualities and could potentially be adapted for inferring effective population sizes, but methods based on continuous random fields or process convolutions can be computationally challenging for large data sets, and some spline methods require selection or modeling of the number and location of knots.
\par
A recent method by \cite{faulk2017} uses shrinkage priors in combination with Markov random fields to perform nonparametric smoothing with locally-adaptive properties.
This is a fully Bayesian method that does not require the use of knots and avoids the costly computations of inverting dense covariance matrices.
Computations instead take advantage of the sparsity in the precision matrix of the Markov random field to avoid matrix inversion.
\cite{faulk2017} compared different specifications of their shrinkage prior Markov random field (SPMRF) models and found that putting a horseshoe prior on the $k$th order differences between successive function values had superior performance when applied to underlying functions with sharp breaks or varying levels of smoothness.
We refer to the model with the horseshoe prior as a horseshoe Markov random field (HSMRF).
\par
In this paper, we propose an adaptation of the HSMRF approach of \cite{faulk2017} for use in phylodynamic inference with coalescent priors.
We devise a new MCMC scheme for the model that uses efficient, tuning-parameter-free, high-dimensional block updates.
We provide an implementation of this MCMC in the  program \texttt{RevBayes}, which allows us to target the joint distribution of genealogy, evolutionary model parameters, and effective population size parameters.
We also develop a method for setting the hyperparameter on the prior for the global shrinkage parameter for coalescent data.
We use simulations to compare the performance of the HSMRF model to that of a GMRF model and show that our model has lower bias and higher precision across a set of population trajectories that are difficult to estimate.
We then apply our model to two real data examples that are well-known in the phylodynamics literature and compare its performance to other popular nonparametric methods.
The first example reanalyzes epidemiological dynamics of hepatitis C virus in Egypt and the second looks at estimation of ancient bison population size changes from DNA data.

\section{Methods}
\label{s:methods}
\subsection{Sequence Data and Substitution Model }
Suppose we have a set of $n$ aligned RNA or DNA sequences for a set of $L$ sites within a gene.
We assume the sequences come from a random sample of $n$ individuals from a well-mixed population, where samples were collected potentially at different times.
Let $\mathbf{Y}$ be the $n \times L$ sequence alignment matrix.
We assume the sites are fully linked with no recombination possible between the sequences.
This allows us to assume the existence of a genealogy $\boldsymbol{g}$, which is a rooted bifurcating tree that describes the ancestral relationships among the sampled individuals.
\par
We assume that $\mathbf{Y}$ is generated by a continuous time Markov chain (CTMC) substitution model that models the evolution of the discrete states (\emph{e.g.}, A,C,T,G for DNA)  along the genealogy $\boldsymbol{g}$ for each alignment site.
A variety of substitution models are available and are typically differentiated by the form of the transition matrix $M(\boldsymbol{\Omega})$, which controls the substitution rates in the CTMC for the nucleotide bases with a set of parameters $\boldsymbol{\Omega}$ (see \cite{yang2014} for examples).
Let the likelihood of the sequence data given the genealogy and substitution parameters be denoted by $p(\mathbf{Y}\mid \boldsymbol{g}, \boldsymbol{\Omega}).$

\subsection{Coalescent}
\label{ss:methods:coal}
Suppose that we now have a genealogy $\boldsymbol{g}$, where branch lengths of the genealogical tree are measured in units of clock time (\emph{e.g.}, years).
To build a Bayesian hierarchical model, we need a prior density for $\boldsymbol{g}$.
The times at which two lineages merge into a common ancestor on the tree are called coalescent times.
The coalescent model provides a probabilistic framework for relating the coalescent times in the sample to the effective size of the population.
\cite{kingman1982} developed the coalescent model for a constant effective population size and \cite{griffiths1994} extended it for varying effective population sizes.
\par
\begin{figure}
	\begin{center}
		\includegraphics[width=\textwidth]{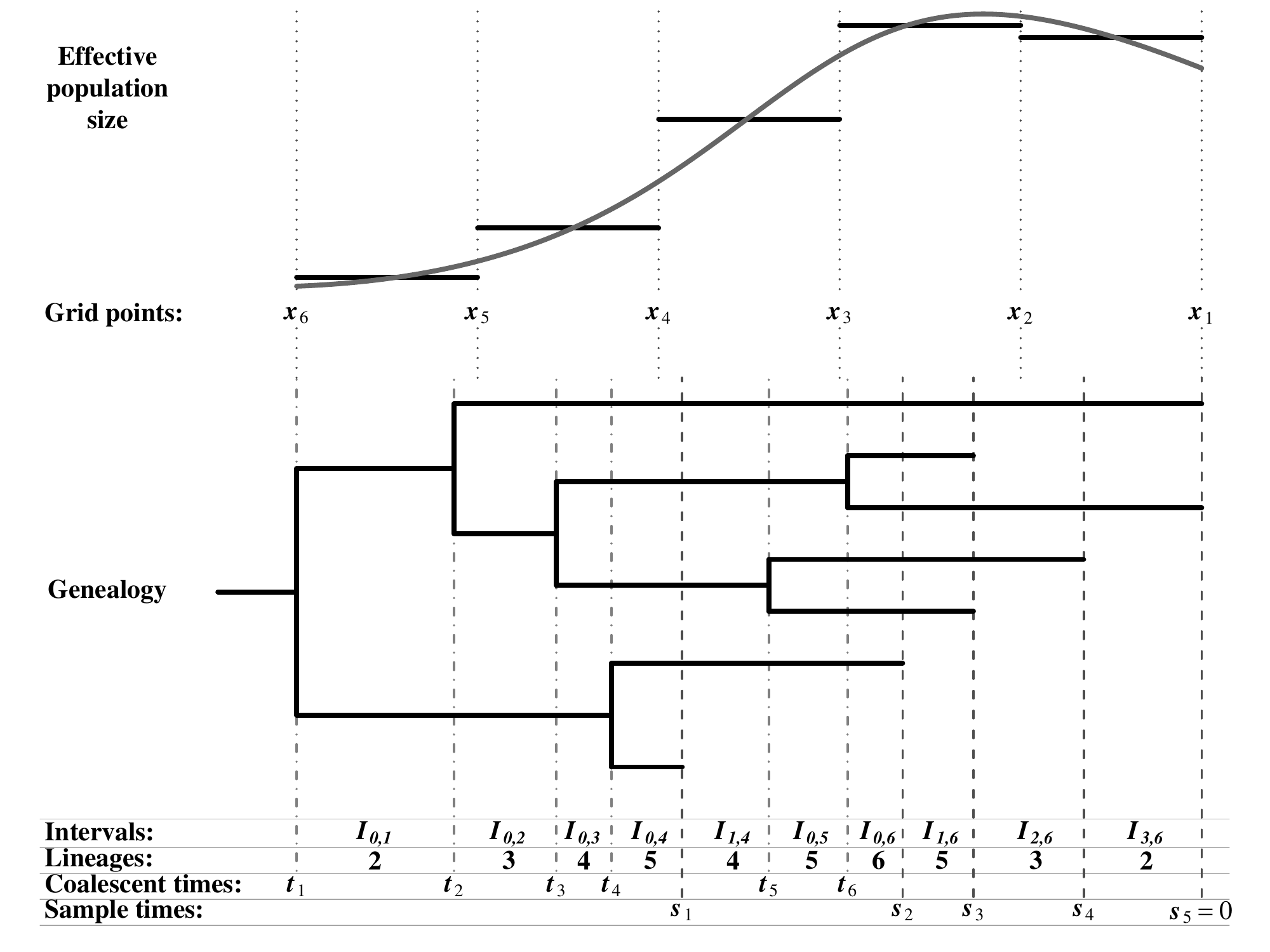}
	\end{center}
	\caption{Effective population size trajectory and associated genealogical tree under heterochronous sampling. The top panel shows a continuous effective population size trajectory (gray) and an associated piecewise constant approximation to it. Also shown are the relationships between the genealogy and sampling times $s_i$, coalescent times $t_i$, intervals $I_{i,k}$, number of lineages $n_{i,k}$, and the uniform grid points, $x_h$, used for approximating coalescent densities.  \label{coalFig}  }
\end{figure}

Let the $n-1$ coalescent times arising from genealogy $\boldsymbol{g}$ be denoted by $0 < t_{n-1} < \dots < t_1$, where 0 is the present and time is measured backward from there.
We will assume the general case where sampling of the genetic sequences occurs at different times (\emph{heterochronous sampling}), which will include the special case where all sampling occurs at time 0 (\emph{isochronous sampling}).
We denote the set of unique sampling times as $s_m = 0 < s_{m-1} < \dots < s_1 < t_1$ for samples of size $n_m, \dots, n_1$, respectively, where $n = \sum_{j=1}^{m}n_j$ and we assume no sample times are equal to coalescent times (Figure \ref{coalFig}).
We let $\boldsymbol{s}$ denote the vector of sampling times.
Further, we let the intervals that end with a coalescent event be denoted $I_{0,k} = (\text{max}\{t_{k+1}, s_j\} , t_k  ], \, \text{for } s_j < t_k \text{and } k=1,\dots, n-1,$ and let the intervals that end with a sampling event be denoted $I_{i,k} = (\text{max}\{t_{k+1}, s_{j+i}\} , s_{j+i-1}  ], \, \text{for } s_{j+i-1} > t_{k+1}\text{ and } s_{j+i} < t_k,\, k=1,\dots, n-1$.
For $k=n-1$, we substitute $t_{k+1}=0$. 
We let $n_{i,k}$ be the number of lineages present in interval $I_{i,k}$ and let the vector of number of lineages be denoted $\boldsymbol{n}$.
Further, we denote the number of unique sampling times in interval $(t_{k+1}, t_k]$ as $m_k$, where $m=1+\sum_{k=1}^{n-1}m_k $. 
The joint density of the coalescent times given $\boldsymbol{s}$ and the effective population size trajectory $N_e(t)$ can then be written as
\begin{equation}
\begin{split}
%\begin{align}
p(t_1, \dots, t_{n-1} \mid \boldsymbol{s},\boldsymbol{n}, N_e(t)) &= \prod_{k=1}^{n-1} p(t_k \mid t_{k+1},  \boldsymbol{s},\boldsymbol{n}, N_e(t))  \\
&=\prod_{k=1}^{n-1} \frac{C_{0,k}}{N_e(t_k)} e^{- \sum_{i=0}^{m_k} \int_{I_{i,k}}\frac{C_{i,k}}{N_e(t)} dt},
\end{split}
\label{coalLik}
\end{equation}
%\end{align}
where $C_{i,k} = {n_{i,k} \choose 2}$ is the coalescent factor \citep{felsen1999}.
This model can be seen as an inhomogeneous Markov point process where the conditional intensity is $C_{i,k} [N_e(t)]^{-1}$  \citep{palacios2013}.
\par
Here we assume $N_e(t)$ is an unknown continuous function, so the integrals in equation (\ref{coalLik}) must be computed with numerical approximation techniques.
We follow \cite{palacios2012}, \cite{gill2013}, and \cite{lan2015} and use discrete approximations of the integrals over a finite grid. 
We construct a regular grid, $\boldsymbol{x}=\{x_h\}_{h=1}^{H+1}$, and set the end points of the grid $\boldsymbol{x}$ such that $x_1 = 0$  and $x_{H+1} = t_1$ (Figure \ref{coalFig}).
This results in $H$ grid cells and $H+1$ cell boundaries.
Now for $t\in (x_h, x_{h+1} ]$, we have $N_e(t)\approx \exp[\theta_h]$, where $\theta_h$ is an unknown model parameter.
This implies that $\boldsymbol{\theta} = \{\theta_h\}_{h=1}^{H}$ is a piecewise-constant approximation to $f(t) = \ln[N_e(t)]$ for $t \in [s_m, t_1]$.
The piecewise constant population size can be integrated analytically, leading to a discrete approximation to the likelihood in equation \ref{coalLik}.
The details of this approximation are provided in Appendix \ref{sectionApprox}.

\subsection{Prior for Effective Population Size Trajectory}
\label{ss:methods:prior}
Next we develop a prior for the unknown function $N_{e}(t)$ that describes the effective population size trajectory over time.
Let $\boldsymbol{\theta} = (\theta_1, \dots, \theta_H )$ be a vector of parameters that govern the effective population size trajectory $N_e(t)$.
We propose using a SPMRF model \citep{faulk2017} for $\boldsymbol{\theta}$, which is a type of Markov model where the $p$th-order differences in the forward-time evolution of the sequence $\left\{\theta_h \right\}_{h=1}^{H}$ are independent and follow a shrinkage prior distribution.
We define the $p$th-order forward difference as $\Delta^p \theta_l \equiv (-1)^{p}\sum_{j=0}^{p}(-1)^j {{p}\choose {j}}\theta_{l + j - p + 1}$, for $l = p,\dots,H-1$, which is a discrete approximation to the $p$th derivative of $f(t)$ evaluated at $t$.
If we assume a horseshoe distribution \citep{carvalho2010horseshoe} as our shrinkage prior on the order-$p$ differences in $\boldsymbol{\theta}$, then
\begin{equation}
  \Delta^p\theta_l \mid \gamma \sim \mathcal{HS}(\gamma),
  \label{hsEqn}
\end{equation}
where the location parameter of the horseshoe distribution is zero and $\gamma$ is the scale parameter and controls how much $f(t)$ is allowed to vary \textit{a priori}.
Following \cite{carvalho2010horseshoe}, we put a half-Cauchy prior on $\gamma$ with scale hyperparameter $\zeta$, so that $\gamma \sim \mathcal{C}^+(0, \zeta)$.
We chose the half-Cauchy here because it has desirable properties as a prior on a scale parameter \citep{gelman2006prior, polson2012half} and its single hyperparameter simplifies implementation.
Depending on the order $p$ of the model, we also place proper priors on $\theta_1,\dots,\theta_p$.
To do this, we start by setting $\theta_1 \sim \mathcal{N}(\mu, \sigma^2)$, where $\mu$ and $\sigma$ are hyperparameters typically set to create a diffuse prior.
Then for $p \geq 2$ and $q = 1, \dots, p-1$, we let $\Delta^q\theta_q\thinvert\gamma \sim \mathcal{HS}(a_q\gamma)$, where $a_q = 2^{-(p-q)/2}$, which follows from the recursive property and independence of the order-$p$ differences.
For example, for $p = 2$,  $a_1 = 2^{-1/2}$, and for $p=3$,  $ a_2 = 2^{-1/2} $ and $a_1=4^{-1/2} $.
We will refer to this specific model formulation as a state-space formulation of a HSMRF.
\par
The horseshoe distribution is leptokurtic with an infinite spike in density at zero and Cauchy-like tails.
In our setting, this combination results in small $\theta$ differences being shrunk toward zero and larger differences being maintained, which corresponds to smoothing over smaller noisy signals while retaining the ability to adapt to rapid functional changes.
This is in contrast to the normal distribution, which has higher density around medium-sized values and normal tails.
These attributes result in noisier estimates and reduced ability to capture abrupt functional changes.
Different shrinkage priors will result in different levels of shrinkage and therefore different smoothing behavior.
\cite{faulk2017} found that the horseshoe prior performed better than the Laplace prior in terms of bias and precision for nonparametric smoothing with SPMRFs, but we do not investigate the effect of different shrinkage priors here.
\par
The horseshoe density does not have a closed form (although see \cite{faulk2017} for an approximation in closed form).
However, a horseshoe distribution can be represented hierarchically as a scale mixture of normal distributions by introducing a latent scale parameter that follows a half-Cauchy distribution \citep{carvalho2010horseshoe}.
That is, if $\tau_l \sim \mathcal{C}^+(0, \gamma)$ and  $\Delta^p\theta_l\mid\tau_l \sim \mathcal{N}(0, \tau_l^2)$, then integrating over $\tau_l$ results in the marginal relationship in equation \ref{hsEqn}.
\par
The hierarchical HSMRF models are a type of $p$th-order normal random walk with separate variance parameters for each increment.
The inherent Markov properties and properties of the normal distribution allow the joint distribution of $\boldsymbol{\theta}$ conditional on the vector of scale parameters $\boldsymbol{\tau}$ to be expressed $p(\boldsymbol{\theta}\mid \boldsymbol{\tau}, \mu, \sigma^2) = p(\theta_1 \mid \mu, \sigma^2)p(\Delta^1\theta_1, \dots,\Delta^p\theta_p,\Delta^p\theta_{p+1},\dots \Delta^p\theta_{H-1} \mid \boldsymbol{\tau} )$, which results in a multivariate normal distribution with mean $\boldsymbol{\mu}$ and precision matrix $\mathbf{Q}(\boldsymbol{\tau})$.
Specifically, $\boldsymbol{\theta}$ follows a Gaussian Markov random field \citep[GMRF;][]{rue2005gaussian} conditional on $\boldsymbol{\tau}$, where the order $p$ of the differencing in $\theta$ determines the structure of the sparse $\mathbf{Q}(\boldsymbol{\tau})$.
For the models presented here, $\boldsymbol{\mu} = \mu\boldsymbol{1}$, where $\mu$ is a constant and $\boldsymbol{1}$ is a vector of ones.
We specify $p(\boldsymbol{\tau})$ by assuming that the $\tau$'s are independent $\mathcal{C}^+(0, \gamma)$-distributed random variables, where $\tau_l \sim \mathcal{C}^+(0, \gamma)$ for $l=p,\dots, H-1$ and $\tau_l \sim \mathcal{C}^+(0, a_l\gamma)$ for $l=1,\dots,p-1$ and $p\geq2$.
The marginal joint distribution of $\boldsymbol{\theta}$ that results from integrating over $\boldsymbol{\tau}$ is a HSMRF.
Note that a GMRF model results when a single scale parameter $\tau$ is used for all order-$p$ differences in $\boldsymbol{\theta}$.
For our GMRF models, we use $\tau \sim \mathcal{C}^+(0, \zeta)$, where $\zeta$ is a fixed hyperparameter.
The order of the HSMRF will determine the amount of smoothing, with higher orders resulting in more smoothing.
We only consider first-order and second-order models here.
In practice, we use the state-space formulation described previously but with the independent hierarchical representations of the horseshoe distributions for the individual order-$p$ differences, which improves computational efficiency over the conditional multivariate normal representation.

\subsection{Posterior Inference}
\label{ss:methods:post}
For the case where we have a fixed genealogical tree, $\boldsymbol{g}$, which consists of sampling times $\boldsymbol{s}$ and coalescent times $\boldsymbol{t}$, the posterior distribution of the parameters $\left\{ \boldsymbol{\theta},\boldsymbol{\tau}, \gamma \right\} $ can be written as
\begin{equation}
  p(\boldsymbol{\theta},\boldsymbol{\tau},\gamma \mid \boldsymbol{g} ) \propto  p(\boldsymbol{g} \mid \boldsymbol{\theta}) p(\boldsymbol{\theta} \mid \boldsymbol{\tau})p(\boldsymbol{\tau}\mid \gamma) p(\gamma).
  \label{fixedPost}
\end{equation}
Here $\boldsymbol{g}$ is considered data and we assume the coalescent times are known.
Then $p(\boldsymbol{g} \mid \boldsymbol{\theta})$ is the coalescent likelihood and $p(\boldsymbol{\theta} \mid \boldsymbol{\tau})p(\boldsymbol{\tau}\mid \gamma) p(\gamma)$ is the HSMRF prior described in Section \ref{ss:methods:prior}.
For our GMRF models, the righthand side of equation \ref{fixedPost} becomes $p(\boldsymbol{g} \mid \boldsymbol{\theta}) p(\boldsymbol{\theta} \mid \tau)p(\tau)$.
\par
For our analyses with fixed genealogical trees, we follow \cite{faulk2017} and \cite{lan2015} and use Hamiltonian Monte Carlo \citep[HMC;][]{neal2011mcmc} for posterior inference.
HMC performs joint proposals for the parameters that are typically far from the current parameter state and have high acceptance rates, resulting in efficient posterior sampling.
We used the \texttt{Stan} computing environment \citep{carpenter2016} for implementing HMC.
Specifically, we used the open source package \texttt{rstan} \citep{rstan-software:2017}, which provides a platform for fitting models using HMC in the \texttt{R} computing environment \citep{rmanual}.
Our \texttt{R} package titled \texttt{spmrf} allows for easy implementation of our models for use on fixed genealogical trees via a wrapper to the \texttt{rstan} tools.
A link to the package code is provided in the  Supporting Information section. 
We present a method for objectively setting the scale hyperparameter $\zeta$ of the prior distribution of the global smoothing parameter $\gamma$ in Appendix \ref{sectionGscale}.
\par
When there are genetic sequence data available and we want to jointly estimate evolutionary parameters, coalescent times, and population size trajectories, our posterior can be written as
\begin{equation}
  p(\boldsymbol{g},\boldsymbol{\Omega},\boldsymbol{\theta},\boldsymbol{\tau},\gamma \mid \mathbf{Y}  ) \propto  p(\mathbf{Y}\mid \boldsymbol{g},\boldsymbol{\Omega})p(\boldsymbol{g} \mid \boldsymbol{\theta}) p(\boldsymbol{\Omega} )  
   p(\boldsymbol{\theta} \mid \boldsymbol{\tau})p(\boldsymbol{\tau}\mid \gamma) p(\gamma) ,
\end{equation}
where $\mathbf{Y}$ are the sequence data and $\boldsymbol{\Omega}$ are the parameters related to the DNA substitution model.
The likelihood of the sequence data given the parameters is $p(\mathbf{Y}\mid \boldsymbol{g},\boldsymbol{\Omega})$, and now $p(\boldsymbol{g} \mid \boldsymbol{\theta})$ is a prior for the genealogy given the population sizes and is proportional to $p(\boldsymbol{g} \mid \boldsymbol{\theta})$ in equation \ref{fixedPost}.
The remaining components are the prior for the evolution parameters $p(\boldsymbol{\Omega} )$ and the HSMRF prior as in equation \ref{fixedPost}.
\par
HMC requires the calculation of gradients over continuous parameter space and therefore cannot be used for inference on discrete parameters.
Therefore, we developed a custom MCMC algorithm that uses a combination of Gibbs sampling, elliptical slice sampling, and the Metropolis-Hastings (MH) algorithm to sample from the joint posterior of the evolution parameters and the effective population size parameters.
In particular, elliptical slice sampling \citep{murray2010} was used to sample from the joint field of log effective population sizes conditional on the latent scale parameters, a Gibbs sampler based on an approach developed by \citep{makalic2016} for horseshoe random variables was used to sample the latent scale parameters conditional on the field parameters, and standard phylogenetic MH steps were used to update the genealogy and substitution model parameters.
We implemented our custom MCMC in \texttt{RevBayes} --- a statistical computing environment geared primarily for phylogenetic inference \citep{hohna2016}.
The standard phylogenetic MH updates mentioned above were already implemented in \texttt{RevBayes}, so we contributed a heterochronous coalescent likelihood calculator, elliptical slice sampling, and Gibbs updates of our model parameters to the \texttt{RevBayes} source code.
The details of the sampling scheme are provided in the Appendix \ref{sectionESS} and a link to the code for implementing our methods for analyzing sequence data is provided in the Supporting Information section. 

\section{Results}
\label{s:results}
\subsection{Simulated Data}
\label{ss:results:sim}
\begin{figure}
	\begin{center}
		\includegraphics[width=\textwidth]{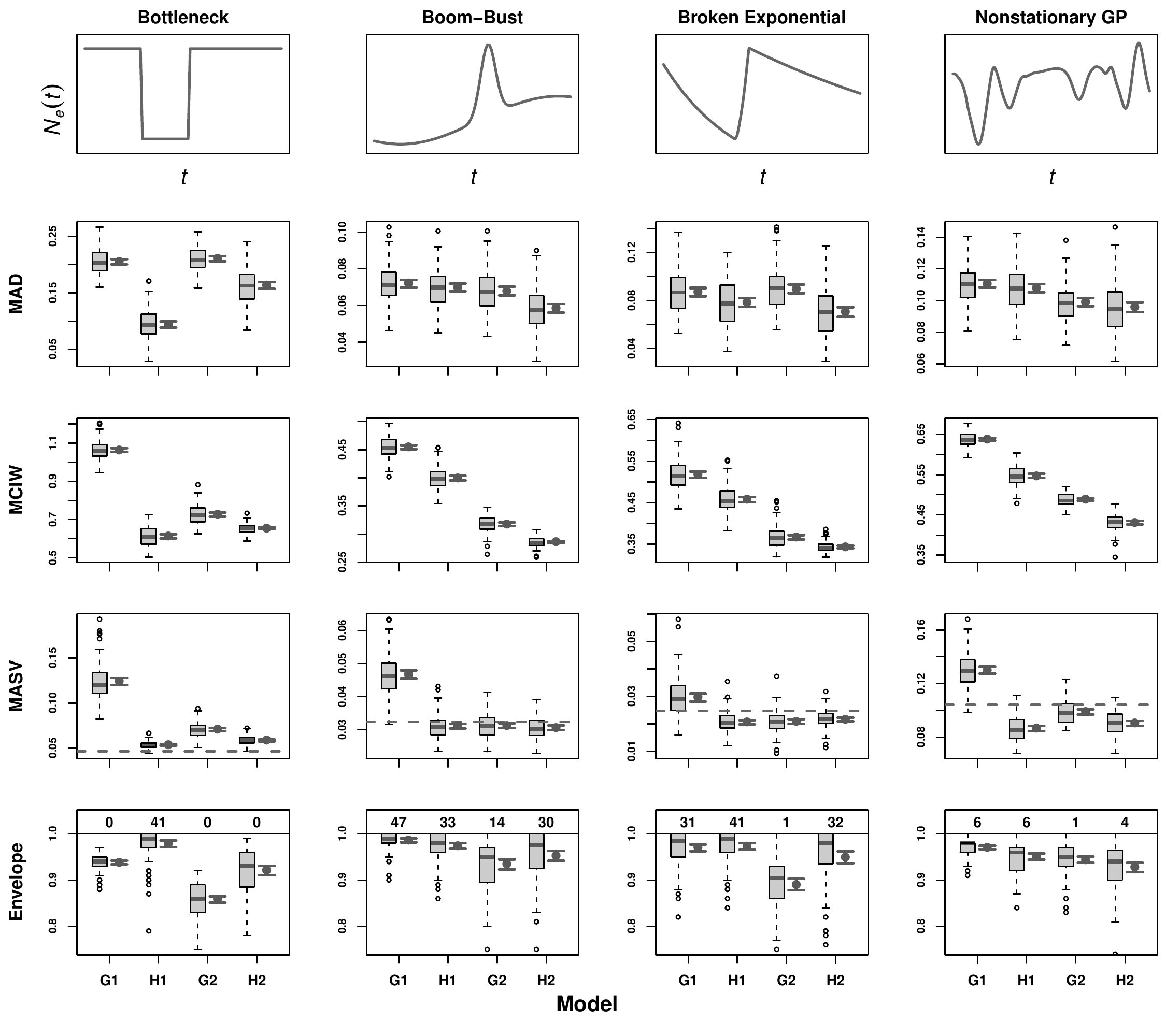}
	\end{center}
	\caption{Effective population size trajectories used in simulations and simulation results by model and scenario.
	Models are GMRF of order 1 (G1) and order 2 (G2) and HSMRF of order 1 (H1) and order 2 (H2).
	Top row shows true effective population size trajectories used to simulate coalescent data.
	Remaining rows show mean absolute deviation (MAD), mean credible interval width (MCIW), mean absolute sequential variation (MASV), and credible interval Envelope.
	Horizontal dashed lines in the third row plots indicate the true mean absolute sequential variation (TMASV) values.
	Shown for each model are standard boxplots of the performance metrics (left) and mean values with 95\% frequentist confidence intervals (right).
	Also shown for Envelope are the number of simulations with Envelope equal to 1.0.\label{simsumFig}  }
\end{figure}

We used simulated data to assess the performance of the HSMRF model relative to the GMRF model.
We investigated four scenarios with different trajectories for $N_e(t)$: (1) Bottleneck (BN), (2) Boom-Bust (BB), (3) Broken Exponential (BE), and (4) Nonstationary Gaussian Process (NGP) realization.
The trajectory shapes are shown at the top of Figure \ref{simsumFig}.
For each scenario, we generated 100 data sets of coalescent times and fit GMRF and HSMRF models of first and second order using the fixed-tree approach.
The scenario descriptions and further methodological details of the simulations are provided in Appendix \ref{apx:sim}.
\begin{figure}
	\begin{center}
		\includegraphics[width=\textwidth]{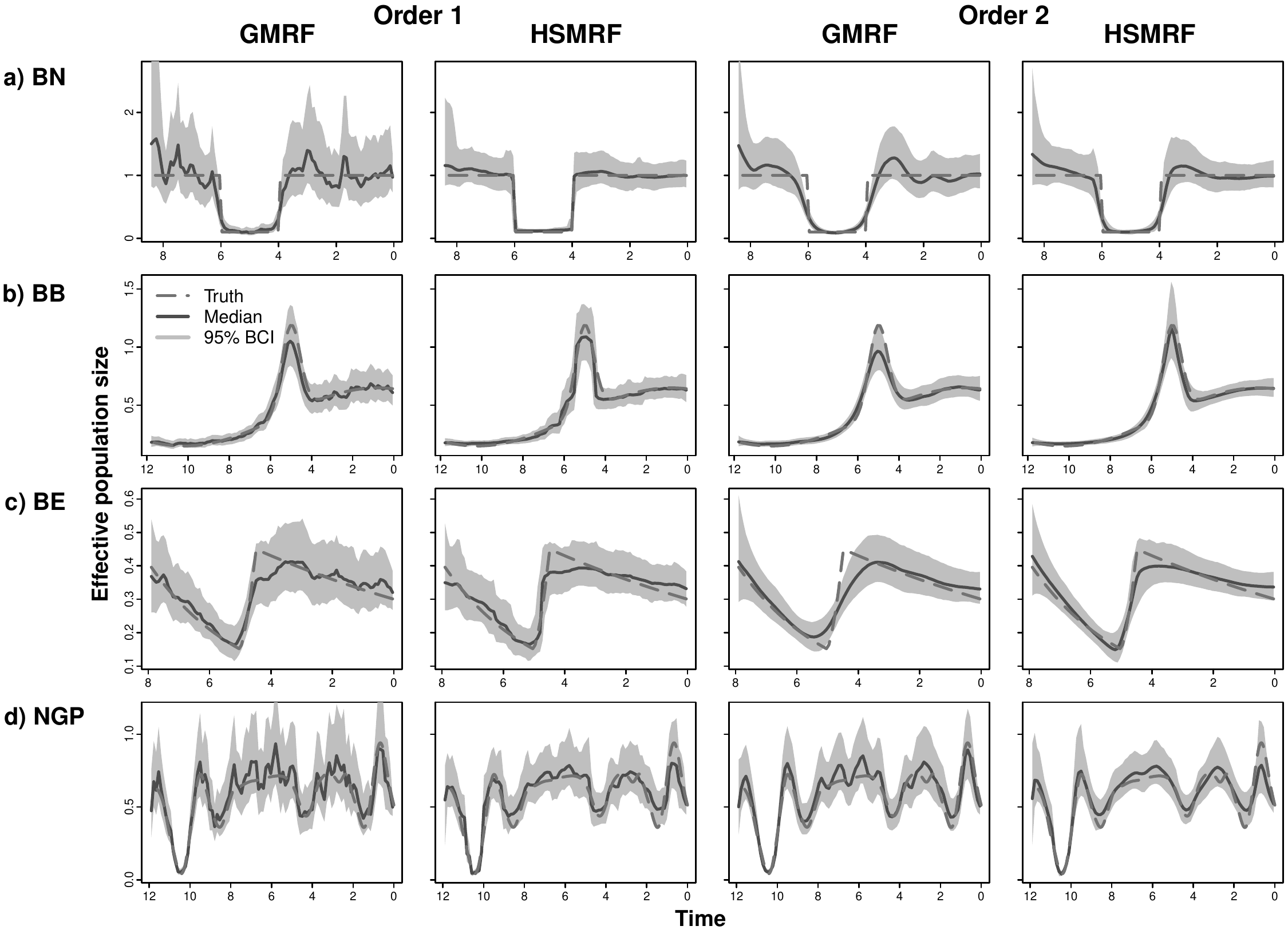}
	\end{center}
	\caption{Example fits of first- and second-order Gaussian Markov random field (GMRF) and horseshoe Markov random field (HSMRF) models for four different simulation scenarios. Scenarios are a) Bottleneck (BN), b) Boom-Bust (BB), c) Broken Exponential (BE), and d) Nonstationary Gaussian Process (NGP). Results for all models within a particular scenario are for the same set of simulated data. Shown are the true effective population size trajectories that generated the data (dashed line), posterior medians of estimated trajectories (solid line) and associated 95\% Bayesian credible intervals (shaded band). \label{simfitsFig}  }
\end{figure}
\par
We assessed the relative performance of the models using a set of summary statistics.
As a measure of bias, we used the mean absolute deviation (MAD) to compare the posterior medians of the trend parameters ($\hat{\theta}_i$) to the true trend values ($\theta_i$):  $\text{MAD} = \frac{1}{H}\sum_{i=1}^{H} \vert \hat{\theta}_i - \theta_i \vert .$
\noindent We assessed the width of the 95\% Bayesian credible intervals (BCIs) using the mean credible interval width (MCIW):
$\text{MCIW} = \frac{1}{H}\sum_{i=1}^{H}\left(  \hat{\theta}_{97.5, i} - \hat{\theta}_{2.5, i} \right)$, where $\hat{\theta}_{97.5, i}$ and $\hat{\theta}_{2.5, i}$ are the 97.5\% and 2.5\% quantiles of the posterior distribution for $\theta_i$.
We assessed the coverage of BCIs using $\text{Envelope}= \frac{1}{H}\sum_{i=1}^{H}I(\theta_i\in [\hat{\theta}_{97.5, i}, \hat{\theta}_{2.5, i}] )$, where $I(\cdot)$ is the indicator function.
To measure local variability in the estimated population trend, we used the mean absolute sequential variation (MASV) of $ \boldsymbol{\hat{\theta} } $, which was computed as $\text{MASV} = \frac{1}{H-1} \sum_{i=1}^{H-1} \vert \hat{\theta}_{i+1} - \hat{\theta}_i \vert .$
We compared the observed MASV to the true MASV (TMASV) in the underlying trend function, which is calculated by substituting true $\theta$'s into the equation for MASV.
For a measure of model complexity, we estimated the effective number of parameters $p_{eff}$ using an approach suggested by \cite{raftery2006}: $p_{eff} = \frac{2}{R-1}\sum_{r=1}^{R}(\mathcal{L}_r - \bar{\mathcal{L}})^2$,
 where $\mathcal{L}_r$ is the log-likelihood evaluated at the parameter values for the $r$th of $R$ samples from the posterior, and $\bar{\mathcal{L}}$ is the mean value of $\mathcal{L}$ across the $R$ samples.
We used the Watanabe-Akaike information criterion \citep[WAIC;][]{watanabe2010} to calculate model weights and rank model performance.
The weight for model $m$ was calculated as $w_m = \exp\left(-0.5\Delta\text{W}_m\right)/\sum_{j=1}^{M}\exp\left(-0.5\Delta\text{W}_j\right)$ for a set of $M$ models, where $\Delta\text{W}_m = \text{WAIC}_m - \min\limits_{j \in M}\text{WAIC}_j$.
We utilized the \texttt{loo} package \citep{vehtari2017} to calculate WAIC.
For a measure of computational efficiency, we calculated the mean effective sample size (ESS) of the posterior samples across parameters for each model and simulated data set and used those with the total sampling times to calculate the mean ESS per second of sampling time.

\begin{table}
\caption{Summary of model selection criteria across 100 simulations by scenario and model set.
WAIC weights were calculated and the best model (greatest WAIC weight) was determined for each simulated data set within each scenario and model set. Metrics shown are the percentage of simulations each model was determined best and the mean model weight across simulations.
Values for each metric are compared among models within each scenario and model set.
Highest percentage of best models is in bold within each scenario and model set.
Scenarios are Bottleneck (BN), Boom-Bust (BB), Broken Exponential (BE), and Nonstationary Gaussian Process (NGP). Models are GMRF of order 1 (G1) and order 2 (G2) and HSMRF of order 1 (H1) and order 2 (H2). } \label{simTabWAIC}
\centering
\begin{tabular}{lllrrrr}
  \toprule
 \textbf{Metric} & \textbf{Model Set} & \textbf{Model}  & \textbf{BN} & \textbf{BB} & \textbf{BE} & \textbf{NGP} \\
\midrule
Best Model (\%) & All Models & G1 &  1 &  9 & 13 &  1 \\
                &            & H1 & \textbf{93} & 14 & 34 &  9 \\
                &            & G2 &  0 &  3 &  1 & 24 \\
                &            & H2 &  6 & \textbf{74} & \textbf{52} & \textbf{66} \\
\cmidrule{2-7}
                & Order 1    & G1 &  1 & \textbf{51} & 29 & 50 \\
                &            & H1 & \textbf{99} & 49 & \textbf{71} & 50 \\
\cmidrule{2-7}
                & Order 2    & G2 &  9 &  9 &  5 & 27 \\
                &            & H2 & \textbf{91} & \textbf{91} & \textbf{95} & \textbf{73} \\
\midrule
Mean Weight       & All Models & G1 & 0.03 & 0.11 & 0.14 & 0.04 \\
                  &            & H1 & 0.89 & 0.15 & 0.35 & 0.09 \\
                  &            & G2 & 0.01 & 0.10 & 0.07 & 0.26 \\
                  &            & H2 & 0.08 & 0.63 & 0.44 & 0.61 \\
\cmidrule{2-7}
                  & Order 1    & G1 & 0.03 & 0.48 & 0.24 & 0.46 \\
                  &            & H1 & 0.97 & 0.52 & 0.76 & 0.54 \\
\cmidrule{2-7}
                  & Order 2    & G2 & 0.12 & 0.11 & 0.11 & 0.43 \\
                  &            & H2 & 0.88 & 0.89 & 0.89 & 0.57 \\
\bottomrule
\end{tabular}
\end{table}

\par For the BN scenario, the HSMRF model clearly had better performance than the GMRF model for the main performance metrics for both model orders (Figure  \ref{simsumFig},  Table \ref{simTabWAIC}, and Table \ref{simTab} in Appendix \ref{apx:sim}).
Example model fits from each scenario provide some intuition for the simulation results (Figure \ref{simfitsFig}).
First order models did better than second order models within model types for the BN scenario.
Differences between model types were not as strong for the other scenarios.
The second-order HSMRF performed the best in terms of MAD, MCIW, and WAIC for the remaining scenarios.
Among second-order models, the HSMRF was clearly favored over the GMRF in terms of WAIC across all scenarios.
However, the HSMRF models were not noticeably different from the second-order GMRF in terms of MASV for the BB and BE scenarios.
The second-order GMRF had mean MASV closer to TMASV than did the second-order HSMRF for the NGP scenario.
Although the GMRF tended to estimate excess variation in the middle section of the trend for the NGP scenario, it did capture the peaks and troughs a little better than the HSMRF in other parts of the trend (see Figure \ref{simfitsFig} for an example).
In all scenarios, the HSMRF had lower $p_{eff}$ compared to the GMRF of the same order.
The GMRF was consistently more computationally efficient than the HSMRF, with mean ESS/sec approximately 1.5 to 6 times higher for models of the same order.
These differences are due to the additional parameters in the HSMRF models.
The second-order models were relatively slow for both model types, but the HSMRF was always slower.
As we show in the following data examples, however, the differences in computational speed between the HSMRF and GMRF models is negligible when genealogies and effective population size trajectories are jointly estimated.

\subsection{Egyptian Hepatitis C Virus}
\label{ss:results:HCV}
The hepatitis C virus (HCV) is a blood-borne RNA virus that exclusively infects humans.
HCV infection is often asymptomatic, but can lead to liver disease and liver failure.
HCV infections have historically had high prevalence in Egypt \citep{miller2010hcv}.
This is thought to be due to past widespread use of unsanitary medical practices in the region.
Of particular interest is a treatment for the parasite disease schistosomiasis known as parenteral antischistosomal therapy (PAT), which uses intravenous injections.
PAT was practiced from the 1920's to 1980's in Egypt and is thought to have contributed to the spread of HCV during that period due to unsterilized injection equipment \citep{frank2000}.

\par
We analyze 63 RNA sequences of type 4 with 411 base pairs from the E1 region of the HCV genome that were collected in 1993 in Egypt \citep{ray2000}.
\cite{pybus2003} used a piecewise demographic model for effective population size with a period of exponential growth between two periods of constant population size and concluded that the HCV population grew exponentially during the period of PAT treatment.
Other authors have applied nonparametric methods to estimate the effective population size trajectory for these data \citep[\emph{e.g.},][]{drummond2005, minin2008sky, palacios2013}.
Different nonparametric methods lead to different estimated trajectories and different levels of uncertainty.
We are interested in estimating the rapid change of HCV effective population size during the epidemic. 
\par
We fit six different nonparametric models to these data: 1) Bayesian Skyline --- a piecewise constant/linear model with estimable locations of change-points \citep[SkyLine;][]{drummond2005}, 2) Bayesian Skyride \citep[SkyRide;][]{minin2008sky}  3) GMRF-1 (similar to Bayesian Skygrid, \cite{gill2013}), 4) GMRF-2, 5) HSMRF-1, and 6) HSMRF-2.
We note that the SkyRide model is also a type of GMRF model where the non-uniform grid cell boundaries are determined by coalescent events.
For all six models we jointly estimated the evolutionary model parameters, genealogies, and effective population size parameters.
We used the program \texttt{BEAST} implementation of the SkyLine and SkyRide models \citep{drummond2012}, and used our own \texttt{RevBayes} implementation of the GMRF and HSMRF models.
Although the Skygrid implementation of the GMRF-1 model is available in \texttt{BEAST}, the GMRF-2 and the HSMRF models are not, so we decided to use common software for the GMRF and HSMRF models.
For the GMRF and HSMRF models we used 100 equally-spaced grid cells where the first 99 ended at a fixed boundary of 227 years before 1993, and the final cell captured any coalescent events beyond the boundary (see Appendix \ref{sectionGrids} for discussion on setting grids).
The SkyLine model requires specification of the number of discrete population intervals, where each interval describes a piecewise constant population size between two coalescent events.
We used 20 population intervals to allow fair flexibility to capture sharp features in the population trajectory.
Further details about the MCMC implementation and computation times are provided in Appendix \ref{sectionImp}.
For model comparison, we calculated posterior model probabilities using marginal likelihood estimates calculated with steppingstone sampling \citep{xie2010}.
See Appendix \ref{sectionPostProb} for details on calculation of posterior model probabilities.

\begin{figure}[!t]
	\begin{center}
		\includegraphics[width=\textwidth]{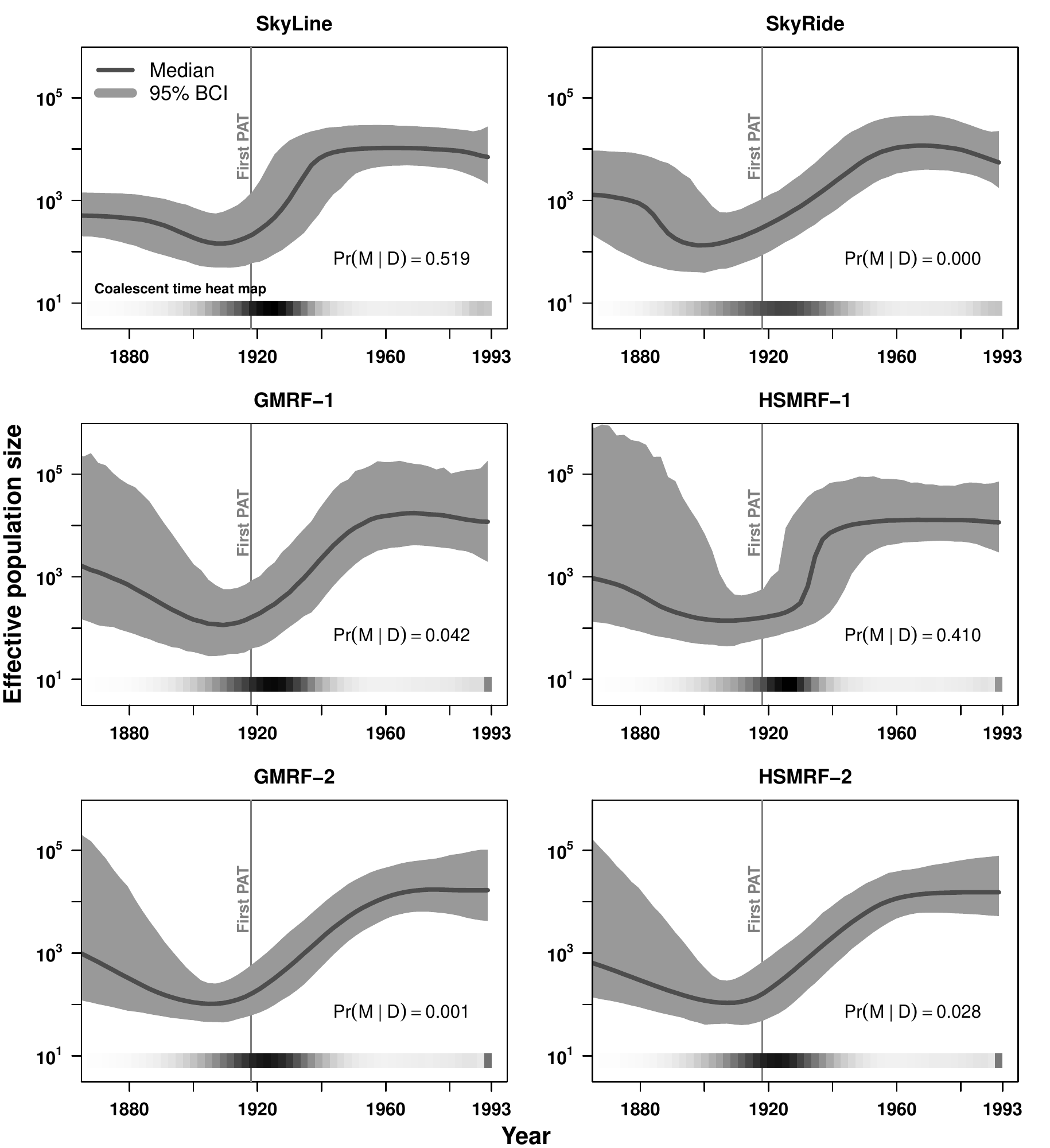}
	\end{center}
	\caption{Posterior medians (solid black lines) of effective population sizes and associated 95\% credible intervals (grey shaded areas) for the HCV data for the Bayesian Skyline (SkyLine), Bayesian Skyride (SkyRide), Gaussian Markov random field of order 1 (GMRF-1) and order 2 (GMRF-2), and horseshoe Markov random field of order 1 (HSMRF-1) and order 2 (HSMRF-2).
	Also shown for each model are posterior model probabilities ($\text{Pr}(\text{M}\mid\text{D})$) and heat maps of mean posterior frequencies of coalescent times.
	A vertical reference line is shown at year 1918, which is the year PAT was introduced. \label{hcvFig}  }
\end{figure}
\par
While the broad pattern of the demographic trajectory was similar among the six models, they differed in the estimated rate of change in effective population size and in the uncertainty around the effective population size estimates (Figure \ref{hcvFig}).
The SkyLine and HSMRF-1 models had the highest posterior model probabilities, with the SkyLine favored a little over the HSMRF-1 (Figure \ref{hcvFig}).
The shape of the median trajectory from the HSMRF-1 model was similar to that of the SkyLine model, yet the HSMRF-1 model showed a very rapid increase in population between 1925 and 1945, while the SkyLine and other models showed more gradual increases that started earlier and ended later.
The increase estimated by the SkyRide model lasted the longest, starting near 1900 and ending near 1970.
The HSMRF and the SkyLine also showed relatively constant population size following the increase in the mid 20th century, while the SkyRide and GMRF-1 models showed a decrease after 1970.
\par
In addition to differing in the rate of population growth after the epidemic began, the models differed in their estimates of when the epidemic began.
The posterior mean densities of frequencies of coalescent times provide an indication of when the HCV epidemic started (Figure \ref{hcvFig}).
The results of the HSMRF-1 support the idea that HCV epidemic started after PAT was introduced and suggest that early PAT campaigns may have used less sanitary practices and contributed more to the spread of HCV than the major PAT campaigns started in the 1950's.
Plots of the effective population trajectories covering the entire span of the coalescent times are provided with further discussion in Appendix \ref{sectionHCVapx}.

\subsection{Beringian Steppe Bison}
\label{ss:results:bison}

Modern molecular methods have allowed the recovery of DNA samples from specimens that lived hundreds to hundreds of thousands of years ago \citep{paabo2004, shapiro2014}.
Large mammals that lived in the Northern Hemisphere during the Pleistocene and Holocene epochs have been a valuable source of this ancient DNA due to conditions favorable for specimen preservation in the northern latitudes \citep[\emph{e.g.,}][]{shapiro2004, lorenzen2011}.   %dropped haile2009
We focus on bison (\emph{Bison} spp.) that lived on the steppe-tundra of Northern Asia and Europe and crossed into North America over the Bering land bridge during the middle to late Pleistocene \citep{froese2017}. 
Interest has been in determining whether human impact or climate and related habitat change instigated the decline of bison across their range during the late Pleistocene.
\cite{shapiro2004} used a parametric piecewise-exponential model for the bison effective population size and estimated that the time of transition from population growth to decline was 37 thousand years ago (kya).
\cite{drummond2005} used the more flexible SkyLine model, which indicated a more rounded and prolonged peak in population size followed by a rapid decline and bottleneck around 10 kya.
Here we use a modified version of the bison data described by \cite{shapiro2004} and fit coalescent models directly to the sequence data as with the HCV data.
We make qualitative comparisons among the resulting estimated population trajectories and in relation to some benchmark times describing the arrival of humans and the period of the Last Glacial Maximum (LGM).
\par
We analyze 152 sequences (135 ancient and 17 modern) of mitrochondrial DNA with 602 base pairs from the mitochondrial control region.
DNA was extracted from bison fossils from Alaska (68), Canada (46), Siberia (13), the lower 48 United States (6), and China (2).
Sample dates were estimated for the ancient samples using radiocarbon dating, with dates ranging up to 59k years.
We treat the calibrated radiocarbon dates as known in the following analyses.
These data are the same as those used by \cite{gill2013}, and are slightly modified from the data first described by \cite{shapiro2004} to remove sequences identified as potentially contaminated with young radiocarbon \citep{shapiro2010} and include additional sequences generated since generation of the initial data set.
In this data set, radiocarbon dates are calibrated to calendar time using the IntCal09 calibration curve \citep{reimer2009}.
\begin{figure}[!t]
	\begin{center}
		\includegraphics[width=\textwidth]{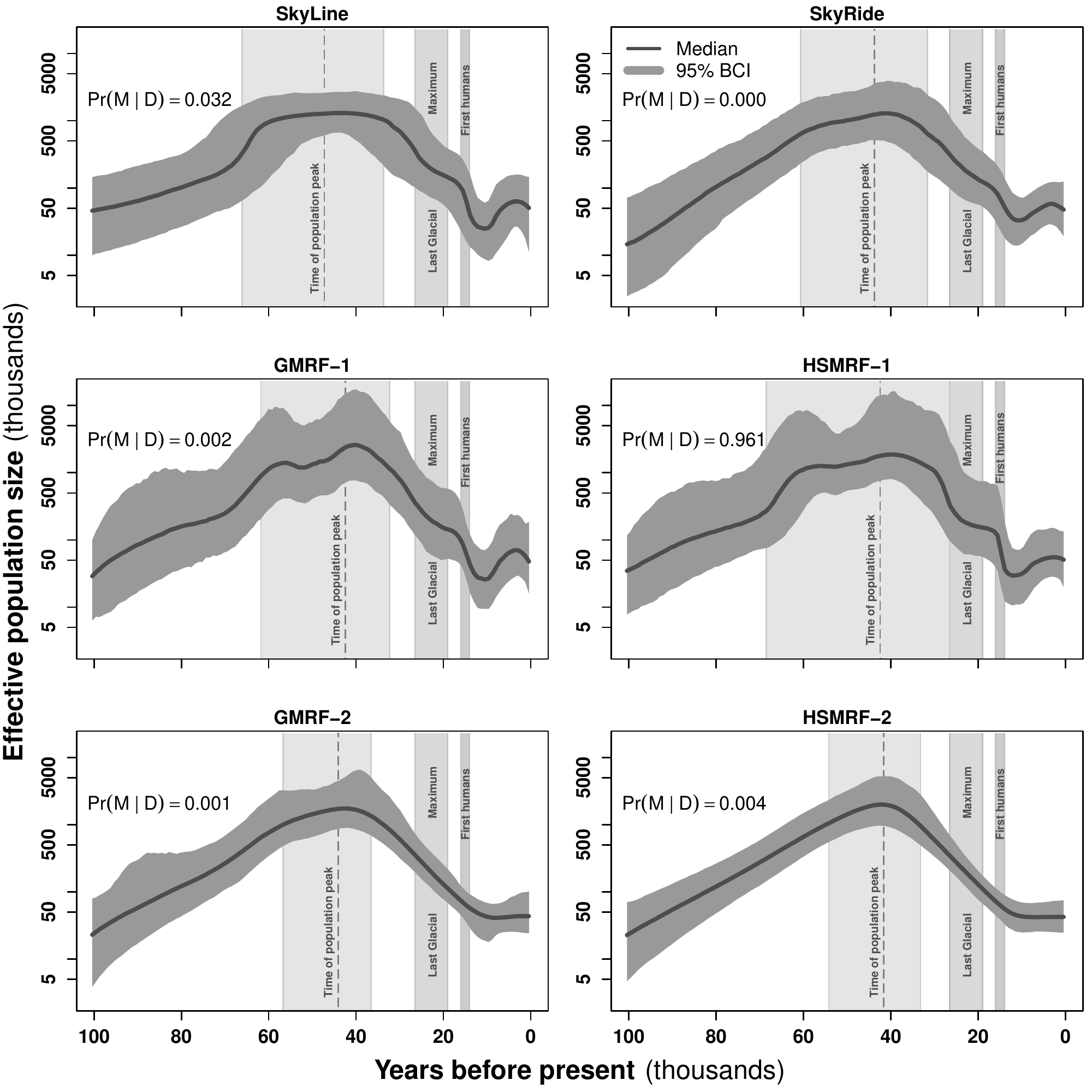}
	\end{center}
	\caption{Posterior medians of effective population sizes and associated 95\% credible intervals obtained from the bison DNA sequence data using the Bayesian Skyline (SkyLine), Bayesian Skyride (SkyRide), and GMRF and HSMRF models of order 1 and order 2.
	Also shown for each model are posterior model probabilities ($\text{Pr}(\text{M}\mid\text{D})$) and posterior median and 95\% credible intervals for the time of peak effective population size.
	The period of the Last Glacial Maximum and timing of first human settlement in North America are shown for reference.
	\label{bisonFigB}  }
\end{figure}
\par
The LGM in the Northern Hemisphere is estimated to have occurred between 26.5 to 19 kya \citep{clark2009}.
A small, isolated population of humans existed in central Beringia, including, potentially, the land bridge that connected the continents during the LGM \citep{llamas2016}.  
Humans may have ventured into eastern Beringia (Alaska and Yukon) as early as 26 kya \citep{bourgeon2017}, but there is as yet no evidence of continuous occupation until 14 kya  \citep{easton2011, holmes2011}.
Humans probably first entered continental North America via a western coastal route that became available close to 16 kya \citep{llamas2016, heintzman2016}, where they would have encountered the population of steppe bison that were isolated in the south with the coalescence of the Laurentide and Cordilleran glaciers \citep{shapiro2004, heintzman2016}.
Because the majority of our bison samples were collected in North America, we used 16-14 kya as the time of first human occupation.
\par
We used methods similar to those used in the HCV example.
We also calculated posterior distributions for the time of the peak in population size.
Method details can be found in Appendices \ref{sectionImp} and  \ref{sectionPostProb}.
\par
While the broad pattern of an increase followed by a decrease in effective population size was recovered by all six models, the timing and nature of the population size change differed considerably between them  (Figure \ref{bisonFigB}).
The HSMRF-1 model had the highest posterior model probability among the six models.
The posterior median trajectory from the HSMRF-1 model was most similar to the SkyLine model, but the credible intervals for the HSMRF-1 model were most similar to the GMRF-1 model.
The second-order models both produced strongly piecewise-linear trajectories with relatively narrow credible intervals, but had low posterior probability and smoothed over some of the local features displayed by other models.
The HSMRF-1 model displayed a more complex descent from the peak size to the present in comparison to the other models, and the areas of rapid descent are coincident with the arrival of humans in eastern Beringia and ice-free North America and the initial retreat of the glaciers, both of which are coincident with changes in habitat.
All models suggested that the overall decline in population size started before the LGM, and all had median time of population peak between 41.6 and 47.3 kya, but uncertainty in the time of peak population size varied widely across the models.

\section{Discussion}

We introduced a novel and fully Bayesian method for nonparametric inference of changes in effective population size that we call the HSMRF.
This method utilizes a shrinkage prior known as the horseshoe distribution, which allows more flexibility to respond to rapid changes in effective population size trajectories, yet also generates smoother trajectories in comparison to standard GMRF methods.
Our simulations demonstrated that the HSMRF had lower bias and higher precision than the GMRF and was able to recover the underlying true trajectories better in most cases.
\par
There are many situations where the local adaptivity of the HSMRF models would provide advantages over the GMRF and other models.
In infectious disease dynamics, examples that could lead to rapid changes in effective population sizes include sudden changes in contact rates due to behavioral changes or quarantine, or sudden changes in the infection rate due to introduction of treatment or vaccine.
At a macro-evolutionary scale, sudden changes in effective population size could be brought on by sudden population collapse (e.g., extinction) or rapid expansions due to dispersals or ecological release.
As we have demonstrated, in situations like these the GMRF and other models tend to smooth over the sharp changes  that the HSMRF can capture.
\par
Our results from both data examples indicated that the properties of the population size trajectories estimated by the HSMRF-1 model were somewhere between those from the GMRF-1 model and the SkyLine model.
The SkyLine model is a type of change-point model, which suggests the HSMRF-1 can produce behavior of change-point models without explicitly needing to specify number or location of change points.
\par
We demonstrated in our simulations that second-order models for either the HSMRF or GMRF formulations can perform better than first-order models in many cases.
Although the second-order models did not perform as well as the first-order models in our particular data examples, they would likely do well in other examples with smoother trajectories.
Among the second-order models, the HSMRF did as well or better than the GMRF for the simulated examples and had higher posterior model probabilities for both of the data examples.
\par
Second-order models have not been used much for estimating effective population sizes previously.
\cite{palacios2013}, whose method assumes a fixed and known genealogy, tested an integrated Brownian motion (IBM) prior for their GP model for the purpose of testing prior sensitivity but did not use the prior beyond that.
The IBM prior is equivalent to the second-order GMRF in continuous time.
Our use of second-order GMRF model for jointly estimating genealogy and effective population size trajectory is the first we are aware of in the literature.
The second-order GMRF and HSMRF can have similar performance in many cases, but HSMRF has the advantage of added flexibility when needed, so it is a reasonable default choice over the GMRF.
We suggest that researchers fit both orders and use a metric such as Bayes factors to select the best order of model for the data.

\section*{Acknowledgments}
J.R.F. and V.N.M. were supported by the NIH grant U54 GM111274.
V.N.M. was supported by the NIH grant R01 AI107034.
A.F.M. was supported by ARCS Foundation Fellowship.
B.S. was supported by the NSF grant DEB-1754461.
We thank the Associate Editor and Reviewers for their constructive criticism and helpful suggestions.

\section*{Supporting Information}
Our \texttt{R} package titled \texttt{spmrf} can be used to fit our models to fixed genealogical trees and is available at {\tt https://github.com/jrfaulkner/spmrf}.
The data and \texttt{RevBayes} code for fitting our models to the molecular sequence data described in Sections \ref{ss:results:HCV} and \ref{ss:results:bison} is available at {\tt https://github.com/jrfaulkner/phylocode}.

\bibliographystyle{biom}
\bibliography{phylospmrfBib}

\appendix
\counterwithin{figure}{section}
\counterwithin{table}{section}
\counterwithin{equation}{section}

\section{Discrete Approximation to Coalescent Likelihood}
\label{sectionApprox}
Here we assume $N_e(t)$ is an unknown continuous function, so the integrals in equation 1 of the main text must be computed with numerical approximation techniques.
We follow \cite{palacios2012}, \cite{gill2013}, and \cite{lan2015} and use discrete approximations of the integrals over a finite grid.
We assume $N_{e}(t) = \exp\left[f(t)\right]$, where $f(t)$ is a function of continuous time.
We approximate $f(t)$ by estimating it at discrete locations on a fixed grid with uniform spacing.
We construct a regular grid, $\boldsymbol{x}=\{x_h\}_{h=1}^{H+1} $, and set the end points of the grid $\boldsymbol{x}$ such that $x_1 = 0$  and $x_{H+1} = t_1$ (see Figure 1 of main text).
This results in $H$ grid cells and $H+1$ cell boundaries.
Now for $t\in (x_h, x_{h+1} ]$, we have $N_e(t)\approx \exp[\theta_h]$, where $\theta_h$ is an unknown model parameter.
This implies that $\boldsymbol{\theta} = \{\theta_h\}_{h=1}^{H}$ is a piecewise-constant approximation to $f(t) = \ln[N_e(t)]$ for $t \in [s_m, t_1]$.
\par
Calculating the likelihood in equation 1 of the main text requires first sorting the combined set of time points  $\left\{\boldsymbol{t},\boldsymbol{s},\boldsymbol{x}\right\}$ and creating a new set of $D = n + m + H - 3$ half-open subintervals $\{I_d^\prime\}_{d=1}^D$, such that for each $d = 1,\dots,D$ there exists an $i,k,\text{and }h$ that satisfy $I_d^\prime = I_{i,k} \cap (x_h,x_{h+1}]$.
Now the integrals in equation 1 of the main text can be approximated by
\begin{equation}
  \int_{I_{i,k}} \frac{C_{i,k}}{N_e(t)}dt \approx \sum_{I_d^\prime \subset I_{i,k}}^{}  \frac{C_{i,k}}{\exp[\theta_h]}\Delta_d,
  \label{intapprox}
\end{equation}
where $\Delta_d$ is the length of the subinterval $I_d^\prime$.
If we introduce an auxiliary variable $z_d$ that takes the value 1 if interval $I_d$ ends with a coalescent event ($I_d^\prime \subseteq I_{0,k}$) and 0 otherwise, then we can use equation (\ref{intapprox}) to write an approximation to the component of the density in equation 1 of the main text associated with interval $(x_h, x_{h+1}]$  as
\begin{equation}
p(\boldsymbol{z}_h\mid \boldsymbol{s},\boldsymbol{n}, N_e(t) )  = \prod_{I_d^\prime \subset (x_h, x_{h+1}]  }^{} \left\{\frac{C_{i,k}}{\exp[\theta_h]} \right\}^{z_d} \exp\left\{-\frac{C_{i,k}}{\exp[\theta_h]}\Delta_d \right\},
\label{ydens}
\end{equation}
where $\boldsymbol{z}_h$ is the vector of $z_d$ values such that $I_d^\prime \subset (x_h, x_{h+1}]$.
An approximation to the complete density in equation 1 of the main text is then the product of the components in equation (\ref{ydens}):
\begin{equation}
p(t_1,\dots,t_{n-1}\mid \boldsymbol{s},\boldsymbol{n}, N_e(t)) \approx \prod_{h=1}^{H}
p(\boldsymbol{z}_h\mid \boldsymbol{s},\boldsymbol{n}, N_e(t) ).
\end{equation}

 \section{Setting the Global Smoothing Hyperparameter}
\label{sectionGscale}
 The global smoothing parameter $\gamma$ controls the variation in the estimated effective population size trajectory.
 It is therefore important to have a way to select the scale hyperparameter $\zeta$ of the prior distribution of the global smoothing parameter that reduces subjectivity.
 We follow a method suggested \cite{sorbye2014} for intrinsic GMRF models and modified by \cite{faulk2017} for SPMRF models for selecting this hyperparameter.
 Let $\boldsymbol{Q}$ be the precision matrix for the Markov random field corresponding to the model of interest (see \cite{faulk2017} for examples), and $\boldsymbol{\Sigma} = \boldsymbol{Q}^{-1}$ be the covariance matrix with diagonal elements $\Sigma_{ii}$.
 The marginal standard deviation of all components of $\boldsymbol{\theta}$ for a fixed value of $\gamma$ is $ \sigma_\gamma(\theta_i) = \gamma \sigma_{\text{ref}}(\boldsymbol{\theta}) $, where  $\sigma_{\text{ref}}(\boldsymbol{\theta})$ is the geometric mean of the individual marginal standard deviations when $\gamma = 1$ \citep{sorbye2014}.
 We want to set an upper bound $U$ on the average marginal standard deviation of $\theta_i$, such that $ \Pr (\sigma_\gamma(\theta_i) > U ) = \alpha $, where $\alpha$ is some small probability (typically 0.01 to 0.05).
 Using the cumulative probability function for a half-Cauchy distribution, we can find a value of $\zeta$ for a given value of $\sigma_{\text{ref}}(\boldsymbol{\theta})$ specific to a model of interest and given common values of $U$ and $\alpha$ by:
 \begin{equation}
 \zeta = \frac{U}{\sigma_{\text{ref}} (\boldsymbol{\theta})  \tan \left( \frac{\pi}{2}(1-\alpha)  \right)} .
  \label{zetafun}
 \end{equation}
 For phylodynamic inference, we set $U$ equal the estimated standard deviation of the log-transformed values the Skyline estimates of population size \citep{pybus2000} based on a fixed genealogy and set of sample times.
 We choose this value of $U$ since we know that the marginal variances of the $\theta$s should not exceed the variance in the log-Skyline estimates, on average.
 For the examples in this paper, we set $\alpha=0.05$ as the probability of the average marginal standard deviation exceeding $U$.

\section{Elliptical Slice within Gibbs Sampler}
\label{sectionESS}
For models based on sequence data, we used a combination of elliptical slice sampling \citep{murray2010} for the latent effective population size parameters and Gibbs sampling for the latent local and global scale parameters.  The Gibbs sampler was based on a modification of the approach derived by \cite{makalic2016} for Gibbs sampling of horseshoe random variables.

\subsection{Model Specifications}
\subsubsection{HSMRF-1}
Using a state-space representation of the HSMRF where $\mu$ is the fixed overall mean and $\sigma^2$ is a fixed variance for $\theta_1$ and $\zeta$ is the fixed hyperparameter on the global scale, following \cite{makalic2016} the first-order HSMRF model conditional on a set of auxiliary variables can be written:
\begin{align*}
  \boldsymbol{y}\mid \boldsymbol{\theta} &\sim \mathcal{L}(\boldsymbol{y}\mid\boldsymbol{\theta}) \\
  \Delta\theta_j &\sim \mathcal{N}(0, \lambda_j^2 \eta^2 \zeta^2)  \quad j =1, \dots, H-1 \\
  \theta_1 &\sim \mathcal{N}(\mu, \sigma^2)\\
  \theta_i &= \theta_1 + \sum_{j=1}^{i-1} \Delta\theta_j  \quad i = 2,\dots, H \\
  \lambda_j^2 \mid \psi_j  &\sim \mathcal{IG}(1/2, 1/\psi_j) \\
  \eta^2 \mid \xi &\sim \mathcal{IG}(1/2, 1/\xi) \\
  \psi_1,\dots,\psi_{H-1}, \xi &\sim \mathcal{IG}(1/2, 1),
 \end{align*}
\noindent where $\boldsymbol{y}$ is the coalescent data, $\mathcal{L}$ is the coalescent density, and $\mathcal{IG}$ is an inverse-gamma distribution.
This formulation implies that $\lambda_j \sim \mathcal{C}^+(0,1)$ and $\eta \sim \mathcal{C}^+(0,1)$.
We translate this to our original model formulation by allowing the global scale parameter $\gamma \sim \mathcal{C}^+(0, \zeta)$, where $\gamma = \eta\zeta$, and the local scale parameters $\tau_j \sim \mathcal{C}^+(0, \gamma)$, where $\tau_j = \lambda_j\gamma = \lambda_j\eta\zeta$.
This implies that $\Delta\theta_j \sim \mathcal{N}(0, \tau_j^2)$, which is our original way of formulating the model.
\subsubsection{HSMRF-2}
The second-order HSMRF model conditional on a set of auxiliary variables can be written:
\begin{align*}
  \boldsymbol{y}\mid \boldsymbol{\theta} &\sim \mathcal{L}(\boldsymbol{y}\mid\boldsymbol{\theta}) \\
  \Delta\theta_1 &\sim \mathcal{N}(0, \frac{1}{2}\lambda_1^2 \eta^2 \zeta^2)   \\
  \Delta^2\theta_j &\sim \mathcal{N}(0, \lambda_j^2 \eta^2 \zeta^2)  \quad j=2, \dots, H-1 \\
  \theta_1 &\sim \mathcal{N}(\mu, \sigma^2)\\
  \theta_2 &= \theta_1 + \Delta\theta_1  \\
  \theta_j &= \Delta^2\theta_{j-1} + 2\theta_{j-1} -\theta_{j-2}   \quad j = 3,\dots,H \\
  \lambda_j^2 \mid \psi_j  &\sim \mathcal{IG}(1/2, 1/\psi_j) \quad j = 1,\dots,H-1  \\
  \eta^2 \mid \xi &\sim \mathcal{IG}(1/2, 1/\xi) \\
  \psi_1,\dots,\psi_{H-1}, \xi &\sim \mathcal{IG}(1/2, 1),
 \end{align*}
\noindent where $\Delta\theta_1 = \theta_2 - \theta_1$, and $\Delta^2\theta_j = \theta_{j+1}-2\theta_j+\theta_{j-1}$ for $j = 2,\dots,H-1$.  
\subsubsection{GMRF-1}
Similar to the HSMRF-1 model above but absent the local scale parameters, the first-order GMRF model can be written:
\begin{align*}
  \boldsymbol{y}\mid \boldsymbol{\theta} &\sim \mathcal{L}(\boldsymbol{y}\mid\boldsymbol{\theta}) \\
  \Delta\theta_j &\sim \mathcal{N}(0, \eta^2 \zeta^2)  \quad j =1, \dots, H-1 \\
  \theta_1 &\sim \mathcal{N}(\mu, \sigma^2)\\
  \theta_i &= \theta_1 + \sum_{j=1}^{i-1} \Delta\theta_j  \quad i = 2,\dots, H \\
  \eta^2 \mid \xi &\sim \mathcal{IG}(1/2, 1/\xi) \\
  \xi &\sim \mathcal{IG}(1/2, 1).
 \end{align*}
\subsubsection{GMRF-2}
The second-order GMRF model can be written:
\begin{align*}
  \boldsymbol{y}\mid \boldsymbol{\theta} &\sim \mathcal{L}(\boldsymbol{y}\mid\boldsymbol{\theta}) \\
  \Delta\theta_1 &\sim \mathcal{N}(0, \frac{1}{2}\eta^2 \zeta^2)   \\
  \Delta^2\theta_j &\sim \mathcal{N}(0, \eta^2 \zeta^2)  \quad j =2, \dots, H-1 \\
  \theta_1 &\sim \mathcal{N}(\mu, \sigma^2)\\
  \theta_2 &= \theta_1 + \Delta\theta_1  \\
  \theta_j &= \Delta^2\theta_{j-1} + 2\theta_{j-1} -\theta_{j-2}   \quad j = 3,\dots,H \\
  \eta^2 \mid \xi &\sim \mathcal{IG}(1/2, 1/\xi) \\
  \xi &\sim \mathcal{IG}(1/2, 1).
 \end{align*}
\noindent where $\Delta\theta_1 = \theta_2 - \theta_1$, and $\Delta^2\theta_j = \theta_{j+1}-2\theta_j+\theta_{j-1}$ for $j = 2,\dots,H-1$.  

\subsection{Full Conditional Distributions}
\subsubsection{HSMRF-1}
First we describe the full conditional distributions of the latent scale and auxiliary variables used in the Gibbs sampler for the first-order HSMRF.
It can be shown that for $j=1,\dots, H-1$, the full conditional distributions are:
\begin{align*}
p(\lambda_j^2\mid \cdot) &\propto \mathcal{IG}\left(1,\,\, \frac{1}{\psi_j}+\frac{\Delta\theta_j^2}{2\eta^2\zeta^2} \right) \\
p(\eta^2 \mid \cdot) &\propto \mathcal{IG}\left(\frac{H}{2}, \,\, \frac{1}{\xi}+\frac{1}{2\zeta^2}\sum_{j=1}^{H-1}\frac{\Delta\theta_j^2}{\lambda_j^2}  \right) \\
p(\psi_j\mid \cdot) &\propto \mathcal{IG}\left(1, \,\, 1+ \frac{1}{\lambda_j^2} \right) \\
p(\xi \mid \cdot) &\propto \mathcal{IG}\left(1, \,\, 1+ \frac{1}{\eta^2} \right)
\end{align*}
\subsubsection{HSMRF-2}
The full conditional distributions of the latent scale and auxiliary variables for the second-order HSMRF are:
\begin{align*}
p(\lambda_1^2\mid \cdot) &\propto \mathcal{IG}\left(1,\,\, \frac{1}{\psi_1}+\frac{\Delta\theta_1^2}{\eta^2\zeta^2} \right) \\
p(\lambda_j^2\mid \cdot) &\propto \mathcal{IG}\left(1,\,\, \frac{1}{\psi_j}+\frac{\Delta^2\theta_j^2}{2\eta^2\zeta^2} \right) \quad j =2, \dots, H-1 \\
p(\eta^2 \mid \cdot) &\propto \mathcal{IG}\left(\frac{H}{2}, \,\, \frac{1}{\xi}+\frac{\Delta\theta_1^2}{\lambda_1^2\zeta^2}+ \frac{1}{2\zeta^2}\sum_{j=2}^{H-1}\frac{\Delta^2\theta_j^2}{\lambda_j^2}  \right) \\
p(\psi_j\mid \cdot) &\propto \mathcal{IG}\left(1, \,\, 1+ \frac{1}{\lambda_j^2} \right) \\
p(\xi \mid \cdot) &\propto \mathcal{IG}\left(1, \,\, 1+ \frac{1}{\eta^2} \right)
\end{align*}

\subsubsection{GMRF-1}
Similarly, the full conditional distributions for the scale and auxiliary variables for the first-order GMRF are:
\begin{align*}
p(\eta^2 \mid \cdot) &\propto \mathcal{IG}\left(\frac{H}{2}, \,\, \frac{1}{\xi}+\frac{1}{2\zeta^2}\sum_{j=1}^{H-1}\Delta\theta_j^2  \right) \\
p(\xi \mid \cdot) &\propto \mathcal{IG}\left(1, \,\, 1+ \frac{1}{\eta^2} \right)
\end{align*}
\subsubsection{GMRF-2}
The full conditional distributions for the scale and auxiliary variables for the second-order GMRF are:
\begin{align*}
p(\eta^2 \mid \cdot) &\propto \mathcal{IG}\left(\frac{H}{2}, \,\, \frac{1}{\xi}+\frac{\Delta\theta_1^2}{\zeta^2}+\frac{1}{2\zeta^2}\sum_{j=2}^{H-1}\Delta^2\theta_j^2  \right) \\
p(\xi \mid \cdot) &\propto \mathcal{IG}\left(1, \,\, 1+ \frac{1}{\eta^2} \right)
\end{align*}

\subsection{Elliptical Slice and Gibbs Sampling}

We follow the algorithm in Figure 2 (pg 543) of \cite{murray2010} for elliptical slice sampling, but with a few modifications.
Suppose the elements of our observation variable $\boldsymbol{y}$ are conditionally independent given a function of underlying latent Gaussian variables $\boldsymbol{\theta} = \boldsymbol{f} + \mu$, where $\boldsymbol{f} \sim \mathcal{N}(0, \boldsymbol{\Sigma}) $ and $\mu$ is a fixed constant.
We denote the likelihood of $\boldsymbol{y}$ conditional on $\boldsymbol{\theta}$ as $\mathcal{L}(\boldsymbol{y}\mid\boldsymbol{\theta})$.
Following \cite{murray2010}, let $\boldsymbol{f}$ be the current state of the zero-centered field parameters on the natural log scale.
The algorithm proceeds by first selecting $\boldsymbol{\nu} \sim \mathcal{N}(0, \boldsymbol{\Sigma})$ and drawing $u \sim \mathcal{U}(0,1)$.
We set the slice value $ s = \ln{u}  +  \ln\mathcal{L}(\boldsymbol{y}\mid\boldsymbol{f}+\mu)$.
We then draw a proposed angle $\alpha \sim \mathcal{U}(0, 2\pi) $, and define a bracket $\left[ \alpha_{min},\alpha_{max}  \right]  = \left[\alpha - 2\pi, \alpha\right]$.
The current proposal is $\boldsymbol{f}' = \boldsymbol{f}\cos{\alpha} + \boldsymbol{\nu}\sin{\alpha} $.
If $\ln\mathcal{L}(\boldsymbol{y}\mid\boldsymbol{f}'+\mu) > s$, then we accept and set $\boldsymbol{f} = \boldsymbol{f}'$.
Otherwise, we shrink the bracket by setting $\alpha_{min} = \alpha$ if $\alpha < 0$ or setting $\alpha_{max} = \alpha$ if $\alpha \geq 0$, and draw a new $\alpha \sim \mathcal{U}(\alpha_{min}, \alpha_{max}) $.
We then calculate a new proposal and keep shrinking the bracket in this manner until the proposal is accepted.
\par
One modification we make to this process is in drawing the initial $\boldsymbol{f}$ and subsequent $\boldsymbol{\nu}$ vectors.
Instead of using the multivariate normal specification, we use the state-space formulation.
To do this, we first draw $\nu_1 \sim \mathcal{N}(0, \sigma^2)$ and then draw $\Delta\nu_j \sim \mathcal{N}(0,\lambda_j^2 \eta^2 \zeta^2 )$ for $j=1,\dots,n-1$ and calculate $\nu_i = \nu_1 + \sum_{j=1}^{i-1} \Delta \nu_j$ for $i = 2,\dots,n$.
Then, prior to evaluating the likelihood, we need to calculate $\boldsymbol{\theta} = \boldsymbol{f} + \mu$.
The likelihood is then $\mathcal{L}(\boldsymbol{\theta}\mid \boldsymbol{y})$.
This approach allows us to sample the variables as multivariate normal with mean zero without needing to use the multivariate normal distribution and costly computations that come with it.
\par
We can specify $\mu$ and $\sigma^2$ using the natural log of the maximum likelihood estimates (MLE) of $N_e(t)$ on a grid, where the coalescent times are obtained from a fixed maximum clade credibility tree, where $\mu$ is the log of the mean of the MLE estimates and $\sigma^2$ is 4 times their variance.
These should provide reasonable hyperparameters that will not result in too diffuse of a sampling distribution.
\par
We use elliptical slice sampling to sample from the field parameters $\boldsymbol{\theta}$ conditional on the latent scale parameters and use Gibbs sampling to update the latent scale parameters conditional on the field and other parameters.
We alternate between these updates until convergence and the desired number of posterior samples are obtained.

\subsection{Checking Validity of Algorithms}
\begin{figure}[th!]
	\begin{center}
		\includegraphics[width=\textwidth]{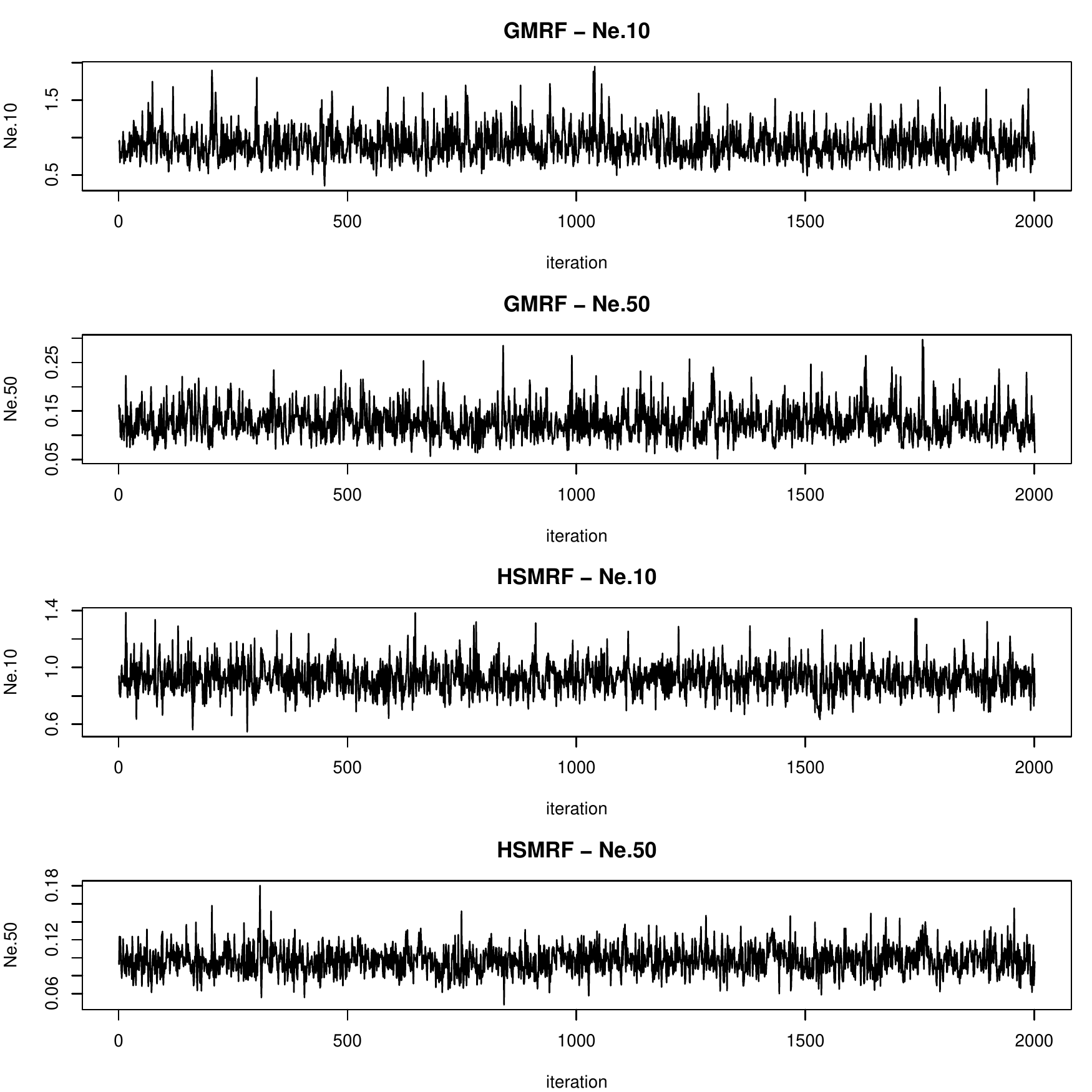}
	\end{center}
	\caption{Trace plots for posterior samples from two $N_e$ parameters from models fit using our elliptical-slice-within-Gibbs sampler in \texttt{RevBayes}.  Examples are for fixed tree coalescent data generated from the Bottleneck scenario used in the main simulations.    }
	\label{tracefig}
\end{figure}

\begin{figure}[th!]
	\begin{center}
		\includegraphics[width=\textwidth]{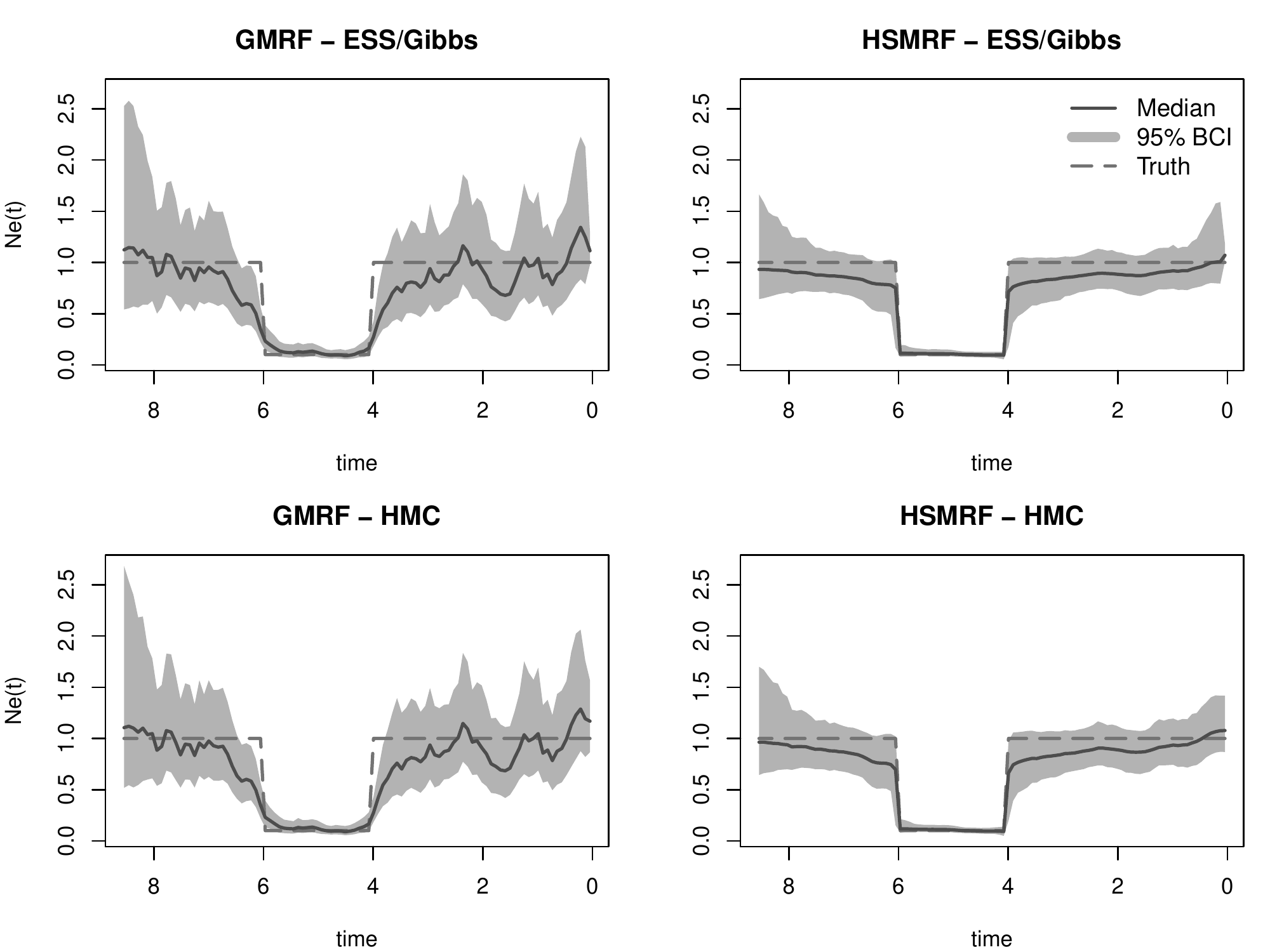}
	\end{center}
	\caption{Posterior medians and 95\% credible intervals for $N_e$ trajectories for two different MCMC samplers.  The top row shows results from the elliptical-slice-within-Gibbs sampler in \texttt{RevBayes}, and the bottom shows results from the HMC sampler in \texttt{Stan} interfaced from the \texttt{spmrf} package. Examples are for fixed tree coalescent data generated from the Bottleneck scenario used in the main simulations. }
	\label{trendfig}
\end{figure}

We performed two checks of our implementation of the random field models in \texttt{RevBayes}.
We simulated coalescent times from the four trajectories that were used in our simulations and generated genealogical trees from those times.
Our first-pass check of our elliptical-slice-within-Gibbs sampler in \texttt{RevBayes} was to feed these trees directly into \texttt{RevBayes} as fixed (as in Section 3.1 of the main text) and compare the results to those obtained with our \texttt{spmrf} package using Hamiltonion Monte Carlo (HMC).
Trace plots for a few parameters from the \texttt{RevBayes} implementation indicate decent mixing (Appendix Figure \ref{tracefig}), and plots of trends from \texttt{RevBayes} implementations do not show appreciable differences from those estimated using HMC with the \texttt{spmrf} package (Appendix Figure \ref{trendfig}).
\par

We then tested our joint inference procedure in \texttt{RevBayes}, estimating the tree topology, coalescent times, and coalescent trajectory.
To reduce computation times, we first down-sampled each tree to 100 tips, ensuring that the retained tips spanned the entire range of non-contemporaneous tips.
We then simulated alignments of 500 sites using mutation rates that produced alignments with an expectation of $\approx 0.93$ substitutions per site.
Thus the simulated alignments were approximately the size of the empirical alignments and contained approximately the same amount of information (number of substitutions).
For simplicity, we employed the Jukes-Cantor substitution model (with no free parameters) with no rate heterogeneity across sites.
When performing the full joint analyses on these datasets, we assumed the clock rate was known (as with the Bison analysis).
All tests indicated that our MCMC sampler was working correctly.
The code we used for conducting these tests is available at {\tt https://github.com/jrfaulkner/phylocode}

%\clearpage

\section{Simulation Details and Additional Results}
\label{apx:sim}

We used simulated data to assess the performance of the HSMRF model relative to the GMRF model.
We investigated four scenarios with different trajectories for $N_e(t)$: (1) Bottleneck (BN), (2) Boom-Bust (BB), (3) Broken Exponential (BE), and (4) Nonstationary Gaussian Process (NGP).
The BN scenario had true $N_e(t) = 0.1$ for $4 \leq t \leq 6$ and $N_e(t) = 1.0$ elsewhere.
The BB scenario had $ N_e(t) = 0.4 + 0.25[\sin((5.5-t)/3) + 0.75\exp(- 2.5(t-5)^2  )]$.
The BE trajectory was $N_e(t) = \exp(-1.20+0.09t)$ for $ 0 \leq t < 4.5$, $N_e(t) = \exp(9.09 -2.20t) $ for $4.5 \leq t < 5$, and $N_e(t) = \exp(-3.57+0.33t)$ for $t \geq 5$.
The trajectory for the NGP scenario, was generated from a Gaussian process with mean 0.55 and a nonstationary Mat\'ern covariance function \citep{paciorek2006}.
The covariance function was constructed so that the length scale increased rapidly in the center of the domain, resulting in a smoother $N_e(t)$ trajectory in the center.
The trajectories used for each scenario are shown at the top of Figure 2 of the main text.
These effective population sizes were set to be small so the the coalescent times would be quick and would mostly fall within a time window specified for each scenario (see below).
\par
For each scenario we generated 100 simulated data sets by first generating a random set of sampling times over a fixed interval and then generating a corresponding random set of coalescent times using the thinning algorithm proposed by \cite{palacios2013} and the true deterministic $N_e(t)$ trajectories defined for each scenario.
For each simulated data set, this is equivalent to assuming we know the fixed genealogical tree for a sample of DNA sequences.
We found that 100 simulations per scenario was sufficient to identify meaningful differences between models without excessive computation time.
We used heterochronous sampling and set the sample sizes based on the complexity of each scenario.
The sample sizes were $n =$ 500, 2,000, 1,000, and 2,000 and the number of lineages sampled at time zero were  $n_m = 50,\,50,\,100,\text{ and }200$ for the BN, BB, BE, and NGP scenarios, respectively.
The remaining sample times followed a uniform distribution on $[0, S]$, where the values of $S$ were 8.0, 11.8, 7.8, and 11.8 for the BN, BB, BE, and NGP scenarios, respectively. We used a fixed grid of 100 cells where the boundary of the 99th cell was $T$ and the final cell collected any coalescent times greater than $T$.
The values of $T$ were 8.37, 11.73, 7.86, and 11.84 for the BN, BB, BE, and NGP scenarios, respectively.
These values were chosen such that the final grid cell contained at least one coalescent time for all of the simulated data sets in a scenario.
For a single fixed tree analysis, we would typically use the final coalescent time as the end of the grid.
However, for the simulations we wanted to keep the grid cells of uniform width across the data sets in a scenario for comparability of results, so we used the same fixed grid for each data set in a scenario.
We chose to use 100 grid cells because that allowed for sufficient resolution to capture features in the underlying trends, and more cells would have increased computation times.
\par
We used HMC to approximate the posterior distribution of model parameters.
For each simulated data set we ran four independent chains, where each chain had 1,000 iterations of adaptation followed by 500 sampling iterations.
This resulted in a total of 2,000 posterior samples.
The hyperparameter on the global scale parameter was selected using the method described in Appendix \ref{sectionGscale} based on the order of the model and the observations from a single data set generated for a scenario.
Summaries of the performance metrics from the simulations are shown in Appendix Table \ref{simTab}.

\begin{table}[!ht]
\caption{Mean values of performance measures across 100 simulations for each model and scenario.
Scenarios are Bottleneck (BN), Boom-Bust (BB), Broken Exponential (BE), and Nonstationary Gaussian Process (NGP).
Models are GMRF of order 1 (G1) and order 2 (G2) and HSMRF of order 1 (H1) and order 2 (H2).
Abbreviations for measures are mean absolute deviation (MAD), mean credible interval width (MCIW), envelope (Env), mean absolute sequential variation (MASV), true MASV (TMASV), effective number of parameters ($p_{eff}$), and mean effective sample size (ESS).
Values of MAD, MCIW, Env, MASV, and TMASV are multiplied by 100 for readability. } \label{simTab}
\centering

\begin{tabular}{llrrrrrrr}
  \toprule
 \textbf{Scenario} & \textbf{Model} & \textbf{MAD} & \textbf{MCIW} & \textbf{Env} & \textbf{MASV} & \textbf{TMASV} & $\boldsymbol{p_{eff}}$  & \textbf{ESS/s}  \\
\midrule
BN & G1 & 20.52  & 106.3 &  93.8 & 12.40  & 4.65 & 55.1 & 8.95 \\
           & H1 &  9.37  &  61.4  & 97.8 &  5.34  & 4.65 & 18.9 & 2.47 \\
           & G2 & 21.06 &   72.6  & 85.8 &  7.05  & 4.65 & 39.4 & 0.78 \\
           & H2 & 16.34 &   65.5  & 92.1 &  5.86  & 4.65 & 34.4 & 0.29 \\
\midrule
BB & G1 & 7.19 & 45.5 & 98.6 & 4.67 & 3.24 & 39.4 & 9.10 \\
          & H1 & 6.98 & 40.0 & 97.4 & 3.11 & 3.24 & 30.6 & 1.25 \\
          & G2 & 6.78 & 31.8 & 93.4 & 3.13 & 3.24 & 25.5 & 0.66 \\
          & H2 & 5.85 & 28.6 & 95.3 & 3.06 & 3.24 & 17.8 & 0.13 \\

\midrule
BE & G1 & 8.70 & 51.7 & 97.0 & 2.96 & 2.47 & 31.0 & 10.22 \\
            & H1 & 7.83 & 45.7 & 97.3 & 2.06 & 2.47 & 21.1 & 2.47 \\
            & G2 & 8.97 & 36.6 & 89.0 & 2.08 & 2.47 & 19.7 & 1.52 \\
            & H2 & 7.04 & 34.3 & 94.9 & 2.16 & 2.47 & 15.2 & 0.53 \\
\midrule
NGP & G1 & 11.06 & 63.8 & 97.0 & 13.00 & 10.42 & 64.7 & 1.72 \\
                 & H1 & 10.78 & 54.7 & 95.1 &  8.65 & 10.42 & 61.4 & 0.54 \\
                 & G2 &  9.90 & 48.8 & 94.4 &  9.89 & 10.42 & 46.9 & 0.07 \\
                 & H2 &  9.58 & 43.1 & 92.8 &  9.04 & 10.42 & 50.4 & 0.05 \\
\bottomrule
\end{tabular}
\end{table}

\section{Guidelines for Constructing Grids}
\label{sectionGrids}
The length of a grid and the number of cells will affect resolution of the estimated effective population size trajectory and its uncertainty, and will also affect computation times.
Here we set some general guidelines that will help with setting up a grid.
\par
The number of grid cells will determine the resolution of detail in the estimated effective population size trajectory.
There have to be enough cells to capture important features in the trajectory, but there should also be enough data to support the number of cells.
As a general rule, we suggest selecting the total number of grid cells, $H$, such that $H = \min\{0.8(n-1), 500\}$, where $n$ is the number of sequences in the data set ($n-1$ is the number of coalescent times). 
This is not a hard and fast rule, but estimates of effective population size tend to behave better when there is at least one coalescent event in a cell or in an adjacent cell.
The upper bound of 500 is arbitrary, but the resolution provided by grid densities greater than 500 is typically not worth the cost of additional computing time.
We should point out that we broke this rule with the HCV example (100 grid cells for 62 coalescent times) because the majority of the coalescent times occurred within the first 50\% of the time domain and we wanted better grid resolution in that first half of the grid to capture details of the population trend.
However, we did follow the rule for the bison example which had 151 coalescent times and we used 120 grid cells, which is approximately 0.8 times 151.
\par
Another important factor in setting up a grid is where to place the boundaries on the final cell.
When performing analyses using coalescent times from a single fixed genealogical tree, the end of the grid is typically set equal to the final coalescent time and all grid cells are equally spaced between time zero and the end of the grid.
However, when analyzing sequence data, the genealogical tree is being estimated, which results in a different set of coalescent times for each MCMC iteration.
Since the width of the grid cells will affect the value of the local and global scale parameters of the random fields used to estimate the effective population sizes, it is necessary to fix the location of the grid boundaries across MCMC iterations.
The location of the boundary $T$ between grid cells $H-1$ and $H$ is important because the uncertainty in the effective population size will be inflated in the final cell if it rarely contains coalescent times across MCMC iterations.
\cite{gill2016} suggested setting $T$ for the Skygrid model so that the final grid cell spans the interval between $T$ and infinity but the remaining grid cells are equally spaced.
Note that the final cell is effectively only as wide as the oldest coalescent time, so the difference in width compared to the other cells should not be that great.
We expand on this idea by formulating a general probability rule for setting $T$ based on data.
We want to find $T$ such that
\begin{equation}
  \Pr(\text{TMRCA} > T) = 1-\alpha_T,
  \label{eqtmrca}
\end{equation} where TMRCA is the time to most recent common ancestor, and also the last coalescent time.
We want $\alpha_T$ to be small, so that there is a high probability that the final grid cell contains the TMRCA in each MCMC iteration.
\par
If a posterior distribution for the TMRCA is available from a previous analysis of the data, then one could use that to find $T$ based on a given $\alpha_T$.
Alternatively, if point estimates and measures of uncertainty for the TMRCA have been published for the data of interest from another study, then those could be used to derive a value for $T$ by a assuming a distribution on the TMRCA, such as a log-normal distribution (see examples below).
In the absence of published estimates, one could use quick frequentist methods such as those proposed by \cite{to2015} or \cite{sagulenko2018} to get initial estimates of the TMRCA based on the sequence data without needing to perform a complete Bayesian analysis.
Note that in some instances after the initial grid setup it may be necessary to adjust the grid after running models and assessing the results to make sure the final cells are adequately capturing the final coalescent times.
\par
For the HCV example, we used the estimated TMRCA of 283 years and associated 95\% credible interval of 246 to 320 years reported by \cite{pybus2003}.
Assuming a log-normal distribution for TMRCA, we estimated the standard deviation of the distribution on the log scale to be $\hat{\sigma} = (\ln(283)-\ln(246))/1.96 = 0.071$ and the mean on the log scale to be $\hat{\mu}=\ln(283) = 5.64$. 
Setting $\alpha_T = 0.001$, the quantile of a log-normal $\mathcal{LN}(\hat{\mu}, \hat{\sigma}^2)$ distribution that satisfies equation (\ref{eqtmrca}) is $T = 227$ years. 
For the bison example, we used the estimated TMRCA of 136 kya and 95\% credible interval of 111 to 164 kya reported by \cite{shapiro2004}.
Following the same procedure as with the HCV example and setting $\alpha_T = 0.001$, the quantile of a $\mathcal{LN}(\hat{\mu}, \hat{\sigma}^2)$ distribution that satisfies equation (\ref{eqtmrca}) is $T = 98.7$ kya. 
We rounded up and set $T = 100$ kya.

\section{Implementation Details for Data Examples}
\label{sectionImp}
For the HCV example, we fixed the mean mutation rate to $7.9 \times 10^{-4} $ substitutions/site/year, which is a value estimated by \cite{pybus2001} and used by others for these data.
We used the HKY nucleotide substitution model \citep{hasegawa1985} with gamma distributed rate heterogeneity and invariant sites \citep{yang1994}.
For each of the models run in \texttt{RevBayes} for the HCV example, we ran four chains each with 1 million iterations of burn-in followed by 25 million iterations of sampling thinned at intervals of 20,000 iterations.
For the SkyLine, SkyRide, and SkyGrid models, each had 1 million burn-in followed by 50 million thinned at every 10,000 iterations.
This resulted in 5,000 posterior samples for each model.
\par
As with the HCV data, for the bison example we used \texttt{BEAST} to fit the SkyLine and SkyRide models and used \texttt{RevBayes} to fit the GMRF and HSMRF models.
We used 15 groups for the SkyLine model to match the approach used by \cite{drummond2005}, which
allowed for sufficient flexibility to fit important change points without introducing computational challenges associated with more groups.
To improve mixing and reduce computation time, we used a strict molecular clock with mutation rate set to $5.38 \times 10^{-7}$ substitutions per year, which was based on initial runs in \texttt{BEAST} where the clock rate was estimated under a Skygrid model for the effective population size trajectory.
We used the HKY nucleotide substitution model with gamma distributed rate heterogeneity.
We used steppingstone sampling \citep{xie2010} to estimate marginal likelihoods for calculating posterior model probabilities (see Appendix \ref{sectionPostProb} for details).
For each of the models run in \texttt{RevBayes} for the bison example, we ran four chains each with 1 million iterations of burn-in followed by 80 million iterations of sampling thinned at intervals of 64,000 iterations.
Each of the models run in \texttt{BEAST} for the bison example had 1 million iterations of burn-in.  The SkyLine, SkyRide, and SkyGrid had 250 million, 100 million, and 100 million iterations of sampling, respectively, each were thinned at 50,000, 20,000, and 20,000, respectively.
These settings for the models in the bison example resulted in a total of 5,000 posterior samples per model.
\par
We used the same distributional forms and parameterizations for priors representing model components in common across the models.
Where possible, we also attempted to use the same proposal distributions and maintain the same relative weighting of different MCMC moves in common to the models across the two software packages.
This corresponds to a somewhat non-standard MCMC scheme in \texttt{RevBayes} where one iteration is equal to one generation. 
In \texttt{RevBayes}, typically move weights are generally chosen such that each MCMC iteration contains a number of generations.
\par
All model runs were performed on a cluster running the Centos 7 Linux operating system with Intel\textregistered{} Xeon\textregistered{} X7550 2.0GHz 64-bit processors and 15.3 GB of RAM.
We used \texttt{BEAST} version 1.10.4 with the \texttt{BEAGLE} library activated.
We used \texttt{RevBayes} version 1.0.11, which includes the modifications made to implement the coalescent models described in the main text.
\par
We calculated the total run times for each of the models by including the time for burn-in and the total time run for the sampling iterations.
For the examples run with multiple chains, we calculated the total time by summing the sampling times of each chain and then adding the average burn-in time across chains.
This combined time is an estimate of time it would have taken to run as single chain with a single burn-in period that would result in the same total number of samples from the combined chains.
The combined times from \texttt{RevBayes} are reported with the times from the single long chains run in \texttt{BEAST} (Appendix Table \ref{timeTab}). 
The run times and ESS/hr differed substantially between the models run in \texttt{BEAST} and the models run in \texttt{RevBayes}, but the run times were almost identical for models of the same order run with \texttt{RevBayes}.
\texttt{RevBayes} was constructed to be a flexible modeling platform, but some of the programming approaches that allowed that flexibility have resulted in long run times for some models.
This problem is known and the \texttt{RevBayes} developers are working to address it.
With the differences between \texttt{BEAST} and \texttt{RevBayes} aside, it is reassuring that the run times for the HSMRF models are not practically different from those of the GMRF models.
Sampling the additional number of parameters in the HSMRF does not add much if any computation time because the Gibbs samplers and elliptical slice samplers are operated on vectors of parameters.
Most of the computation time in \texttt{RevBayes} appears to be dominated by sampling the genealogical trees and evolution parameters.
If run times cannot be improved in \texttt{RevBayes}, we will look to implement our HSMRF models in \texttt{BEAST}.
Either way, whether implemented in \texttt{BEAST} or in an optimized version of \texttt{RevBayes}, we expect that the difference in run times between the GMRF and HSMRF models will remain minimal.
\begin{table}[!ht]
\caption{Run times (hrs) , mean effective sample sizes (ESS) for the log effective population size parameters, and mean number of effective samples per hour (ES/hr) for the HCV and bison data examples.}
\label{timeTab}
\centering
\begin{tabular}{lllrrr}
  \toprule
 \textbf{Example} & \textbf{Software} & \textbf{Model} & \textbf{Time} &  \textbf{ESS} & \textbf{ES/hr} \\
\midrule
HCV & BEAST             &  SkyLine   &   2.7  & 1,853.2   & 681.3    \\
       &                & SkyRide    &   3.4  & 4,469.2  &  1,308.7   \\
        &                & SkyGrid   &   3.6  & 2,441.1  &  668.8   \\
        & RevBayes        & GMRF-1   &   522.9 & 1,307.5 & 2.5   \\
        &                & HSMRF-1 &    524.1 & 1,073.5  & 2.0  \\
        &                & GMRF-2   &    521.3  &   468.6  & 0.9  \\
        &                & HSMRF-2 &    521.0  &    457.2 &  0.9 \\
 \midrule
Bison & BEAST              &  SkyLine   &     16.8   & 1,596.9   & 94.9 \\
         &                & SkyRide    &      9.1    & 3,983.5   & 436.3 \\
        &                & SkyGrid   &        9.4    & 3,398.8   & 353.3 \\
        & RevBayes       & GMRF-1   &    339.0       & 2,833.5   & 8.4  \\
        &                & HSMRF-1 &    341.5        & 2,517.4   & 7.4  \\
        &                & GMRF-2   &    338.0       & 2,282.3   & 6.8  \\
        &                & HSMRF-2 &     337.9       & 2,758.1   & 8.2 \\
\bottomrule
\end{tabular}
\end{table}

\section{Calculating Posterior Model Probabilities}
\label{sectionPostProb}
Calculation of posterior model probabilities requires estimates of marginal (or integrated) likelihood values. 
For both the HCV and bison data examples, we used steppingstone sampling \citep{xie2010} to estimate marginal likelihoods.
Steppingstone sampling is a type of path sampling algorithm \citep{gelman1998} which is similar to the thermodynamic integration methods developed by \cite{lartillot2006} and \cite{friel2008} but incorporates importance sampling to improve computational efficiency. 
\par
The stepping stone and thermodynamic methods evaluate the posterior where the likelihood is raised to a power over a sequence of power values between 0.0 and 1.0, which spans the set of densities between the posterior and the prior.
For each model in each data example, we used 50 stones, where the stones represent a set of quantiles of a $\text{Beta}(\alpha, 1.0)$ distribution which are used as the sequence of power values.
We set $\alpha=0.2$ in the beta distribution.
We used the same total number of iterations used in the main analyses but spread equally among the stones.
\begin{table}[!t]
\caption{Summary of model selection results for data examples. Shown are natural log of  marginal likelihood values (logML), natural log of Bayes factors (logBF), Bayes factors (BF), and posterior model probabilities (Pr(M\textbar D)) by model for the HCV and bison data examples.
Bayes factors and posterior model probabilities are calculated relative to the HSMRF-1 model. }\label{mlTab}
\centering
\begin{tabular}{llrrrr}
  \toprule
 \textbf{Example} & \textbf{Model} & \textbf{logML} & \textbf{logBF}  & \textbf{BF} & \textbf{Pr(M\textbar D)}   \\
\midrule
HCV &  SkyLine   &  -6,396.02  &  0.23  &  	1.2644 &  	0.5188
\\
        & SkyRide    &  -6,424.75 &  -28.49  &  	0.0000	 &  0.0000
\\
        & GMRF-1   &   -6,398.54 &  	-2.29  &  	0.1017	 &  0.0417
\\
        & HSMRF-1 &   -6,396.26 &  	 0.00  &  	1.0000 &  	0.4103
\\
        & GMRF-2   &   -6,402.83 &  	-6.57	&  0.0014 &  	0.0006
\\
        & HSMRF-2 &   -6,398.93 &  	-2.67	&  0.0695 &  	0.0285 \\
 \midrule
Bison    &  SkyLine    &  -3,717.53   &  -3.39	  &  0.0336	  &  0.0322
\\
            &  SkyRide    &  -3,731.68	 &  -17.54	  &  0.0000	  &  0.0000
\\
            &  GMRF-1   &  -3,720.27   &  -6.13	  &  0.0022	  &  0.0021
\\
           &  HSMRF-1 &   -3,714.14  &	 0.00	  &  1.0000	  &  0.9611
\\
          &  GMRF-2   &   -3,721.26  &  -7.12	  &  0.0008	  &  0.0008
\\
           &  HSMRF-2 &   -3,719.66  &	 -5.52	  &  0.0040	  &  0.0038 \\
\bottomrule
\end{tabular}
\end{table}

\par
Once we have marginal likelihood estimates, we can calculate Bayes factors \citep{kass1995} and posterior model probabilities to compare evidence for different models.
The posterior odds of Model 1 ($\mathcal{M}_1$) relative to Model 2  ($\mathcal{M}_2$) conditional on the data ($\mathcal{D}$) is calculated as
  \begin{equation*}
\frac{\Pr(\mathcal{M}_1\mid \mathcal{D})}  {\Pr(\mathcal{M}_2 \mid \mathcal{D} )} = \frac{ \Pr(\mathcal{D}\mid \mathcal{M}_1  ) } { \Pr(\mathcal{D}\mid \mathcal{M}_2  ) }   \frac{ \Pr(\mathcal{M}_1)  }   {\Pr(\mathcal {M}_2 ) },
  \end{equation*}
where $\Pr(\mathcal{D} \mid \cdot) $ is the marginal likelihood of the data given a particular model, and the Bayes factor is the ratio of marginal likelihoods: $B_{12} = \Pr(\mathcal{D}\mid \mathcal{M}_1  )/\Pr(\mathcal{D}\mid \mathcal{M}_2 )$.
For our set of six models $\mathcal{M}_1,  \dots, \mathcal{M}_6 $, we calculated the posterior probability of $\mathcal{M}_k$ as 
\begin{equation*}
  \Pr(\mathcal{M}_k \mid \mathcal{D}) = \alpha_k B_{k1}  / \sum_{r=1}^{6}\alpha_r B_{r1}
\end{equation*}
where $\alpha_k = \Pr(\mathcal{M}_k)/\Pr(\mathcal{M}_1)$ is the prior odds of $\mathcal{M}_k$ relative to $\mathcal{M}_1$ (HSMRF-1), and $B_{11}, \dots, B_{61}$ are the Bayes factors calculated relative to $\mathcal{M}_1$.
We assumed equal prior model probabilities, so $B_{11} = \alpha_1 = 1$.
Appendix Table \ref{mlTab} shows results for marginal likelihoods, Bayes factors, and posterior model probabilities from the HCV and bison data examples used in the main text.

\section{Additional Results for HCV Example}
\label{sectionHCVapx}
In the main text for the HCV example, we focused on the period of rapid increase in the effective population size trajectory that generally occurred after the year 1900.
Here we report results for the entire time domain.
\begin{figure}[ht]
	\begin{center}
		\includegraphics[width=\textwidth]{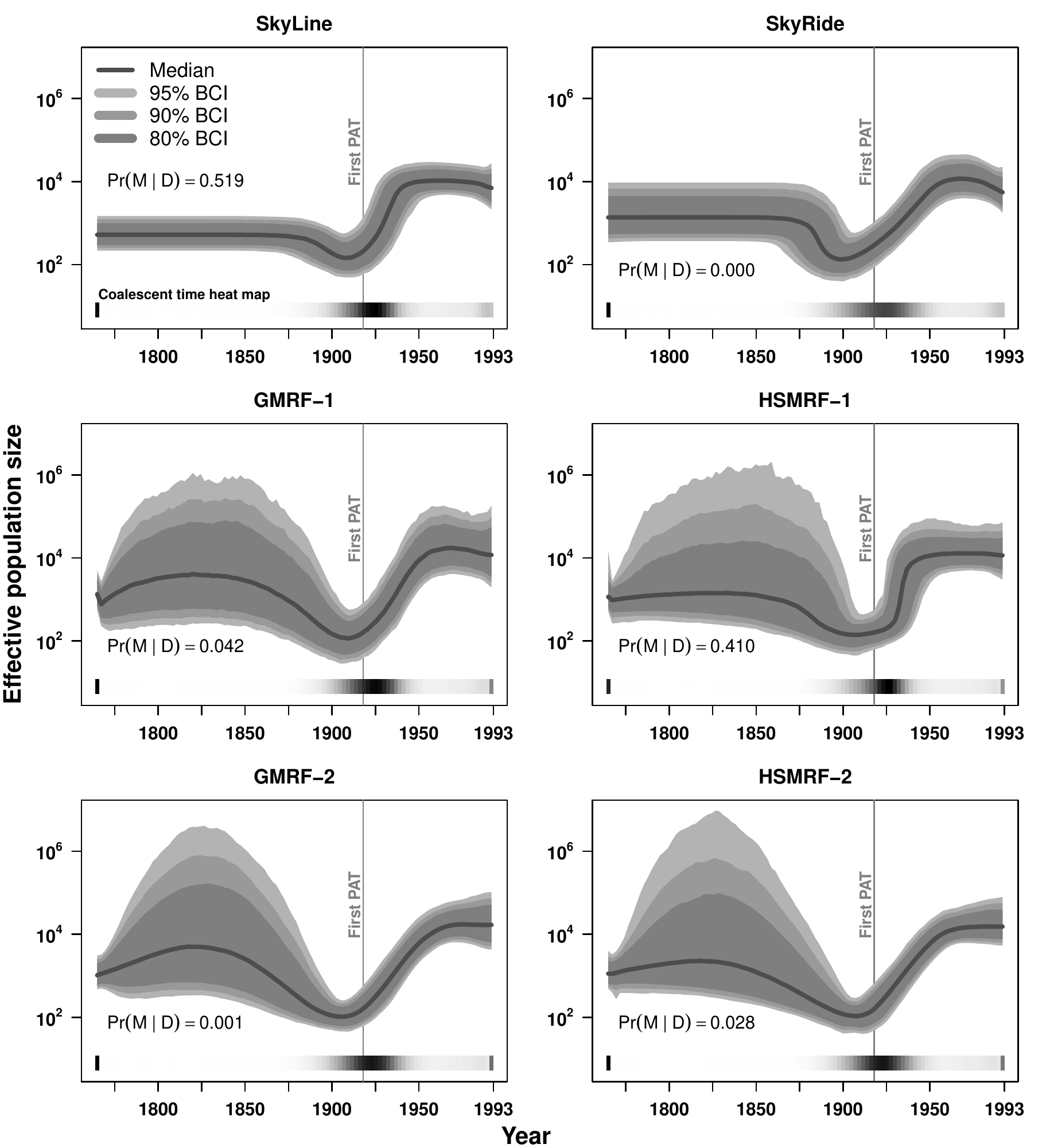}
	\end{center}
	\caption{Posterior medians (solid black lines) of effective population sizes and associated 95\%, 90\%, and 80\% credible intervals (layered gray shaded areas) for the HCV data for the complete time domain for the Bayesian Skyline (SkyLine), Bayesian Skyride (SkyRide), Gaussian Markov random field of order 1 (GMRF-1) and order 2 (GMRF-2), and horseshoe Markov random field of order 1 (HSMRF-1) and order 2 (HSMRF-2).
	Also shown for each model are posterior model probabilities ($\text{Pr}(\text{M}\mid\text{D})$) and heat maps of mean posterior frequencies of coalescent times.
	A vertical reference line is shown at year 1918, which is the year PAT was introduced. \label{hcvfig2}  }
\end{figure}
The heatmaps of the posterior frequencies of coalescent event times show that coalescent events were very unlikely between approximately 1770 and 1870 (Appendix Figure \ref{hcvfig2}) .
The GMRF and HSMRF models had similar levels of uncertainty in the population size trajectories during that gap in coalescent events, but the posterior distributions for effective population sizes were more skewed for the HSMRF models than for the GMRF models (Appendix Figure \ref{hcvfig2}).
The result of this is that the 95\% credible intervals were wider for the HSMRF models compared to the GMRF models, but the corresponding 90\% and 80\% credible intervals were narrower for the HSMRF models compared their GMRF counterparts.
Both the SkyLine and SkyRide models had narrow credible intervals during the gap in coalescent times, with the SkyRide model showing more uncertainty than the SkyLine.
Both of these models use piecewise constant trajectories between coalescent events, which means they are restricted to be nearly flat over the long period without coalescent events.
In contrast, the GMRF and HSMRF models had a large number of grid cells covering the gap in coalescent events and the effective population size was allowed to be different in each grid cell.
The uncertainty in effective population sizes during the gap in coalescent times was likely grossly underestimated for the SkyLine and SkyRide due to the constrained nature of those models.
\par
The GMRF and HSMRF models show a small change the population size trajectory and associated credible intervals in the final grid cell.
This is because the final grid cell extends from year 1766 to infinity and likely capture all of the final four coalescent events for many of the posterior draws.
The change from zero to four coalescent events in the final grid cell is the likely explanation for the abrupt change in estimated effective population size seen on the plots.

\clearpage

\end{document}